\documentclass[fleqn,usenatbib]{mnras}

\usepackage{newtxtext,newtxmath}

\usepackage[T1]{fontenc}

\DeclareRobustCommand{\VAN}[3]{#2}
\let\VANthebibliography\thebibliography
\def\thebibliography{\DeclareRobustCommand{\VAN}[3]{##3}\VANthebibliography}

\usepackage{graphicx}
\usepackage{amsmath}
\usepackage{xcolor}
\usepackage{pdflscape}

\newcommand{\unit}[1]{~\mathrm{#1}}

\newcommand{\e}[1]{\times 10^{#1}}
\newcommand{\numrange}[2]{#1$ to $#2}
\newcommand{\degree}{^\circ}

\newcommand{\Rej}{R_{\rm ej}}
\newcommand{\Mej}{M_{\rm ej}}

\newcommand{\Ekej}{E_{\rm kin,ej}}
\newcommand{\Ethej}{E_{\rm th,ej}}
\newcommand{\Ekejtild}{\tilde{E}_{\rm kin,ej}}

\newcommand{\pvec}{\mathbf{R}}
\newcommand{\gvec}{\mathbf{g}}

\newcommand{\Leng}{L_{\rm eng}}
\newcommand{\Eeng}{E_{\rm eng}}

\newcommand{\Eengtild}{\tilde{E}_{\mathrm{eng}}}
\newcommand{\teng}{t_{\rm eng}}
\newcommand{\Meng}{M_{\rm eng}}

\newcommand{\ttild}{\tilde{t}}
\newcommand{\Menc}{M_{\rm enc}}

\newcommand{\oxygen}{^{16}\mathrm{O}}
\newcommand{\nickel}{^{56}\mathrm{Ni}}
\newcommand{\helium}{^{4}\mathrm{He}}

\newcommand{\castro}{\textsc{Castro}}
\newcommand{\sedona}{\textsc{Sedona}}

\newcommand{\plotone}[1]{\includegraphics[width=\linewidth]{#1}}

\newcommand{\plottwo}[2]{\includegraphics[width=0.45\linewidth]{#1}\includegraphics[width=0.45\linewidth]{#2}}

\definecolor{darkblue}{rgb}{0,0,0.35}
\newcommand{\markchange}[1]{{#1}}

\title[Engine-powered supernovae]{Dynamics and observational signatures of core-collapse supernovae with central 
engines: hydrodynamics simulations with Monte Carlo post-processing}

\author[K.\ Eiden and D.\ Kasen]{
Kiran Eiden$^{1}$\thanks{E-mail: kiran\_eiden@berkeley.edu} and Daniel Kasen$^{1\markchange{,2,3}}$
\\
$^{1}$Department of Astronomy, University of California, Berkeley, CA 94720, USA\\
$^{2}$Theoretical Astrophysics Center, University of California, Berkeley, CA 94720, USA\\
$^{3}$Nuclear Science Division, Lawrence Berkeley National Laboratory, 1 Cyclotron Road, Berkeley, CA 94720, USA
}

\date{Accepted XXX. Received YYY; in original form ZZZ}

\pubyear{2026}

\begin{document}
\label{firstpage}
\pagerange{\pageref{firstpage}--\pageref{lastpage}}
\maketitle

\begin{abstract}
A long-lived central engine embedded in expanding supernova ejecta can alter the dynamics
and observational signatures of the event, producing an unusually luminous, energetic,
and/or rapidly-evolving transient. We use two-dimensional hydrodynamics simulations
to study the effect of a central energy source, varying the amount, rate, and isotropy of
the energy deposition. We post-process the results with a time-dependent Monte Carlo radiation
transport code to extract observational signatures. The engine excavates a bubble
at the centre of the ejecta, which becomes Rayleigh-Taylor unstable. Sufficiently powerful
engines are able to break through the edge of the bubble and accelerate, shred, and compositionally
mix the entire ejecta. The breakout of the engine-driven wind occurs at distinct rupture points,
and the outflowing high-velocity gas may eventually give rise to radio emission. The dynamical
impact of the engine leads to faster rising optical light curves, with photon escape facilitated
by the faster expansion of the ejecta and the opening of low-density channels. For models with
strong engines, the spectra are initially hot and featureless, but later evolve to
resemble those of broad-line Ic supernovae. Under certain conditions, line emission
from ionized, low-velocity material near the centre of the ejecta may be able to escape and produce
narrow emission similar to that seen in interacting supernovae. We discuss how variability in the
engine energy reservoir and injection rate could give rise to a heterogeneous set of events spanning
multiple observational classes, including the fast blue optical transients, broad-line Ic supernovae,
and superluminous supernovae.
\end{abstract}




\section{Introduction}

In recent decades, transient surveys have unveiled a diverse landscape of stellar explosions potentially linked to 
core-collapse supernovae (CCSNe), the luminous events signalling the gravitational collapse of massive stars. Models 
of stellar explosions powered by the radioactive decay of $\nickel$ can explain standard CCSNe, but cannot account 
for many of the unusually energetic or rapidly evolving transients that inhabit this broader 
landscape  \citep{Kasen2017}. Among these peculiar events are superluminous supernovae (SLSNe), broad-line Type Ic supernovae (SNe Ic-BL), and fast blue optical transients (FBOTs). SLSNe (\citealt{Chomiuk2011,Quimby2011,Gal-Yam2012}; see \citealt{Gal-Yam2019,Chen2021,Nicholl2021} for recent reviews) are 10-100 times brighter than ordinary CCSNe. SNe Ic-BL (see, e.g., \citealt{Iwamoto1998,Nomoto2001}) have broad absorption lines in their spectra, indicative of large expansion velocities, and have also been associated with long gamma-ray bursts (GRBs; 
\citealt{Modjaz2011,Cano2017}). FBOTs \citep{Drout2014,Margutti2019,Ho2022} are characterized by a rapid rise to 
peak luminosity spanning $\lesssim 10\unit{d}$ and bluer emission than typical CCSNe, and can achieve peak 
luminosities greater than or comparable to those of SLSNe. Also of interest are some peculiar Type Ib supernovae, 
such as the double-peaked SN 2005bf (see \citealt{Maeda2007}) and the highly energetic ($E_k \sim 
10^{52}\unit{erg}$) SN 2012au, which has late time properties reminiscent of some Type I SLSNe and SNe Ic-BL 
\citep{Milisavljevic2013,Milisavljevic2018}.

There are a variety of proposed mechanisms for generating these transients, and some classes transients may be 
produced by several of these mechanisms or a combination of them. One possibility is that these are CCSNe with 
energy injection from a central compact object, which may take the form of a rapidly-rotating magnetar 
\citep{Usov1992,Thompson1994,Wheeler2000,Maeda2007,Kasen2010,Woosley2010,Metzger2011,Metzger2015} or accreting
black hole or neutron star \citep{Woosley1993,MacFadyen1999,Woosley2012,Dexter2013,Kashiyama2015}. Luminous
transients with accretion-powered engines may also arise from tidal disruption events or merger/common envelope events
involving a star and a compact object \citep{Perley2019,Soker2019,Kuin2019,Kremer2021,Metzger2022,Grichener2025,Tsuna2025},
or ``failed" supernovae that produce a black hole and accretion disc, with energy injection from the disc unbinding
the stellar envelope \citep{Margutti2019,Perley2019,Quataert2019,Antoni2022,Antoni2023}.

Another potentially relevant mechanism is the interaction of the supernova ejecta or outflows with a dense
circumstellar medium (CSM), which under certain conditions can efficiently convert the supernova kinetic energy
to radiation and produce luminous and rapidly evolving light curves \citep{Chevalier2011,Ginzburg2014,Jiang2020,Suzuki2020,Suzuki2021a,Khatami2024,Hamidani2025}. Forming this dense
CSM would require a mass-loss event (or series of events) preceding the explosion that ejects a portion of the
envelope. Such an episode could be the result of wave-driven outbursts from unstable nuclear burning 
\citep{Arnett2011a,Arnett2011b,Quataert2012b,Smith2014,Fuller2017,Fuller2018,Wu2021,Wu2022a}, binary interaction 
with a companion star \citep{Tauris2013,Tauris2015,Ouchi2017,Wu2022b}, or some other mechanism.

In very massive stars, the pair-production instability can lead to contraction of the stellar core and explosive 
nuclear burning. In pair-production supernovae (PPSN, \citealt{Barkat1967,Ober1983,Glatzel1985,Heger2002}) this 
explosive burning completely unbinds the star (mass $\sim \numrange{130}{260}\unit{M_\odot}$), potentially 
producing large quantities of $\nickel$, while in pulsational pair-instability supernovae (PPISNe; 
\citealt{Heger2007,Chatzopolous2012b,Chen2014,Yoshida2016})  material is ejected in a series of pulses, with 
collisions between these shells potentially producing a luminous transient \citep{Heger2002,Heger2007}.

While some Type II SLSNe exhibit narrow hydrogen Balmer lines in their spectra, characteristic of interaction with 
a relatively slow-moving CSM, hydrogen-poor (Type I) SLSNe tend to lack these features. Analysis of X-ray data from 
some Type I SLSNe also appears to disfavour interaction as the primary energy source \citep{Margutti2018}, lending 
support to the theory that a subset of SLSNe are powered by central engines. SNe Ic-BL and associated GRBs tend to 
have large kinetic energies of $\gtrsim 10^{52}\unit{erg}$, and theoretical models of GRBs require some asymmetry 
in the form of relativistic jets or outflows, favouring engine models. Broadband monitoring of the luminous FBOT AT2018cow
revealed an embedded X-ray source with both hard ($\geq 10\unit{keV}$) and soft components, as well as radio emission
consistent with a near-relativistic blast wave \citep{Margutti2019}. Although the FBOT landscape is 
heterogeneous and central engine models may not be able to account for the spectral and photometric properties of 
all FBOTs, some form of engine model could potentially explain at least some fraction of FBOT-like events.

Analytic/semi-analytic models 
\citep{Maeda2007,Woosley2010,Suwa2015,Kasen2010,Dexter2013,Kasen2016,Omand2024,Omand2025} and 1D simulations 
\citep{Kasen2010,Dexter2013,Dessart2018a,Dessart2018b,Dessart2019,Orellana2018,Moriya2022} of CCSNe with 
central engines are able to reproduce some observational properties of these events. These have been fit to 
observational data from SLSNe, SNe Ic-BL, and FBOTs to obtain parameter estimates for engine models (see, e.g., 
\citealt{Nicholl2017b,Moriya2018,Liu2022,Omand2024,Gomez2024, Konyves-Toth2025}). However, certain aspects of the 
dynamics and evolution of engine-powered CCSNe cannot be fully captured by 1D or one-zone models. The engine may 
produce aspherical outflows, e.g., in the form of jets or disc winds. Furthermore, as seen in 1D models 
\citep{Kasen2010}, the wind from the engine excavates a low-density cavity in the centre of the ejecta. In two or more 
dimensions, the thin shell encasing this cavity becomes unstable (see, e.g., 
\citealt{Chevalier1977,Chevalier1992,Jun1998,Bucciantini2004,Gelfand2009} for a discussion of this in the context of 
pulsar wind nebulae, and \citealt{Arons2003} for some discussion of magnetars embedded in CCSN envelopes). The 
instability could fragment the shell and permit the engine-driven wind to break through into the outer ejecta, 
accelerating the ejecta and altering the final structure.

\citet{Couch2011,Papish2014a,Papish2014b}, \citet{Chen2017}, \citet{Barnes2018}, and \citet{Suzuki2022} have 
performed multidimensional hydrodynamics simulations of jet-driven CCSNe, with \citet{Chen2017} also exploring 
wind-like injection and combinations of winds and jets. \citet{Dupont2023} simulated equatorial outflows from 
magnetar-like engines, and found that sufficiently collimated outflows can break out of the progenitor at 
ultrarelativistic velocities. \citet{Chen2016,Chen2020}, \citet{Suzuki2017,Suzuki2019,Suzuki2021b}, and 
\citet{Blondin2017} used multidimensional hydrodynamics simulations to examine the effects of a magnetar-like 
central engine that injects energy isotropically on CCSN ejecta. The simulations were performed in both 2D 
\citep{Chen2016,Suzuki2017,Suzuki2021b,Blondin2017} and 3D (\citealt{Suzuki2019,Chen2020}, again 
\citealt{Blondin2017}), with \citet{Suzuki2017,Suzuki2019} including special relativistic effects and 
\citet{Suzuki2021b} using radiation-hydrodynamics. Broadly, their results indicate that the growth of instabilities 
drives mixing of elements in the ejecta, and that the wind from a sufficiently energetic engine can break apart the 
shell at the edge of the central cavity and flow into the outer ejecta. \citet{Suzuki2018} also calculated light 
curves and broad-band spectral energy distributions based on the results presented in \citet{Suzuki2017}. They 
concluded that magnetar-powered CCSNe are capable of producing bright non-thermal radio and X-ray emission, although the luminosity of the emission depends on the ejecta density structure.

In this study, we perform 2D hydrodynamics simulations of expanding supernova ejecta with a central energy source. 
We expand on previous multidimensional studies\markchange{,} exploring the impact of engines with a range of energy injection 
rates, energy reservoirs, and injection morphologies. We also post-process the simulation results using Monte Carlo 
radiation transport to obtain approximate light curves and spectra, and comment on how these compare to observed 
transients. In Section \ref{sec:bg_and_analytics} we describe the equations used to model the expanding ejecta and the 
central engine, and outline some analytical expectations for the evolution of our models. We discuss our numerical 
simulation setup \markchange{in} Section \ref{sec:simulation_setup}, and present the results of the hydrodynamics simulations in 
Section \ref{sec:results}. We discuss the post-processing setup in Section \ref{sec:observables}, and present some 
tentative conclusions regarding the observational signatures of these events. Finally, we conclude the paper in Section 
\ref{sec:discussion}.

\section{Background: Dynamics of Engine-Driven Bubbles}\label{sec:bg_and_analytics}
    
Here we present the equations we use to model the ejecta structure and energy
injection from the central engine, and discuss the expected dynamical evolution
in 1D. We discuss how these equations are implemented in our hydrodynamical
simulations in Section \ref{sec:simulation_setup} and how our results
compare to these analytic expectations in Section \ref{sec:results}.

\subsection{Initial Ejecta Structure}

We initialize the mass distribution in the ejecta according to the broken power law
of \citet{Chevalier1989a}:
\begin{equation}\label{eq:init_dens_prof}
    \rho(R) = \left\{\begin{array}{cc}
        \rho_t \left(\frac{R}{R_t}\right)^{-d}; & R < R_t\\
        \rho_t \left(\frac{R}{R_t}\right)^{-n}; & R_t \leq R \leq \Rej
    \end{array}\right\},
\end{equation}
where $R$ is the Euclidean distance from the origin, $R_t$ is the radius corresponding
to the ``break" or transition in the power law, $\rho_t$ is the density at the transition
point, $d < 3$ and $n > 5$ are the power law indices, and $\Rej$ is the radius at which
the ejecta meets the surrounding medium. $\rho_t$ in terms of $\Mej$ and $R_t$ is
\begin{equation}\label{eq:tr_dens}
    \rho_t =  \frac{\zeta_\rho}{4 \pi} \left( 
    \frac{\Mej}{R_t^3} \right),
\end{equation}
with
\begin{equation}\label{eq:zeta_rho_def}
    \zeta_\rho \equiv \left[ \frac{1}{3 - d} + \frac{1 - \left(R_t/\Rej\right)^{n-3}}{n - 3} \right]^{-1}.
\end{equation}
The ejecta is assumed to be expanding homologously, so the velocity profile for $R \leq \Rej$ is
\begin{equation}\label{eq:init_vel_prof}
    v(R) = v_t \left(\frac{R}{R_t}\right),
\end{equation}
where $v_t$ is the velocity at the broken power law transition point.

The ejecta kinetic energy $\Ekej$ and the ejecta mass are related by
\begin{equation}\label{eq:E_to_M_ratio}
    \Ekej = \zeta_E \Mej v_t^2,
\end{equation}
where the constant
\begin{equation}\label{eq:zeta_E_def}
    \zeta_E \equiv \left[\frac{1}{2} \frac{(n-3)(3-d)}{(n-5)(5-d)}\right] \left[\frac{(n-d) - 
        (5-d)(R_t/\Rej)^{n-5}}{(n-d) - (3-d)(R_t/\Rej)^{n-3}}\right]
\end{equation}
is set by the density structure.

\subsection{Central Engine Behaviour}

We consider the general case of a central engine that injects energy into the ejecta at a 
rate
\begin{equation}\label{eq:lum_eng}
    \Leng(t) = \frac{\Eeng}{\teng} \frac{k-1}{\left(t/\teng + 1 \right)^{k}},
\end{equation}
where $\Eeng$ is the total energy reservoir and $\teng$ is some characteristic time-scale 
for the engine. For constant $\teng$, the total energy emitted up to a time $t$ is then simply
\begin{equation}\label{eq:E_emit_eng}
    E_{\rm emit}(t) = \Eeng \left[1 - \left(t/\teng + 1\right)^{1 - 
        k}\right].
\end{equation}

In the case of a magnetar central engine approximated as a rotating point dipole in a vaccuum, we have $k = 
2$. The initial energy reservoir is the magnetar rotational energy given by
\begin{equation}\label{eq:init_rot_energy}
    \Eeng = \frac{2 \pi^2 I}{P_0^2} \approx 2\e{52}~\frac{I_{45}}{P_{0,{\rm ms}}^2}\unit{erg},
\end{equation}
where $I$ is the magnetar moment of inertia and $P_0$ is the initial period \markchange{($I_{45}$ and $P_{0,{\rm ms}}$ are these quantities in units of $10^{45}\unit{g~cm^2}$ and $1\unit{ms}$ respectively)}. The characteristic 
time-scale is the magnetar spin-down time-scale; from the Larmor formula, we can derive
\begin{align}
    \teng &= \frac{3 c^3 I}{16 \pi^2} P_0^2 (B R^3 \sin\alpha)^{-2} \nonumber \\&\approx 2\e{3}~I_{45} P_{0,{\rm ms}}^2 (2 B_{15} 
    R_6^3 \sin\alpha)^{-2}\unit{s}.
\end{align}
The additional parameters here are the magnetic field strength $B$ \markchange{(where $B_{15}$ is $B$ in units of $10^{15}\unit{G}$)}, the magnetar radius $R$ \markchange{($R_6$ is $R$ in units of $10^{6}\unit{cm}$)}, and the angle $\alpha$ between the magnetic and rotation axes. We can see that for the default parameter values and $\sin\alpha = \frac{1}{2}$, periods in the range $\numrange{1}{10}\unit{ms}$ yield magnetar energies from $\sim \numrange{10^{50}}{10^{53}\unit{erg}}$ and time-scales on the order of minutes to days.

In the case of an accretion-powered engine, we write the energy injection rate as
\begin{equation}\label{eq:accretion_lum}
    \Leng(t) = \eta c^2 \dot{M}_{\rm acc} \approx \frac{\eta c^2 M_{\rm acc}}{\teng} \frac{k-1}{\left(t/\teng + 1 
    \right)^{k}},
\end{equation}
where $\eta$ is the rest-mass energy conversion efficiency, $M_{\rm acc}$ is an approximation
of the accreted mass, and $\dot{M}_{\rm acc}$ the accretion rate. We assume fallback
accretion, and also assume that the bound stellar material is radially symmetric with density profile
$\rho(r) = \rho_{\rm acc} (r/r_{\rm acc})^{-m}$, where $r_{\rm acc}$ is the outermost initial radius
of the accreted material. The characteristic time-scale (of order the free-fall time) is 
given by
\begin{equation}
    \teng \approx \frac{\pi r_{\rm acc}^{3/2}}{\sqrt{2GM_{\rm enc}(r_{\rm acc})}} \approx 1~\left(\frac{r_{\rm 
    acc}}{R_\odot}\right)^{3/2} \left(\frac{\Menc}{M_\odot}\right)^{-1/2}\unit{hr},
\end{equation}
where $G$ is the gravitational constant, and $\Menc(r_{\rm acc}) = 4\pi \rho_{\rm acc} r_{\rm acc}^3 / (3-m)$
is the enclosed mass at radius $r_{\rm acc}$ \citep{Quataert2012a}. Assuming a shallow density profile with $0 < m 
< 3$, $\dot{M}_{\rm acc}$ for strongly bound material with velocity much less than the escape velocity takes the 
form (\citealt{Dexter2013}; similar to \citealt{Quataert2012a} equation 2)
\begin{equation}
    \dot{M}_{\rm acc} = \frac{2 (3 - m)}{d} \frac{\Menc(r_{\rm acc})}{\teng} \left(\frac{t}{\teng}\right)^{6/m - 3}.
\end{equation}
In the case of marginally bound material, we have
\begin{equation}
    \dot{M}_{\rm acc} = \frac{2 (3 - m)}{3} \frac{\Menc(r_{\rm acc})}{\teng} \left(\frac{t}{\teng}\right)^{-5/3},
\end{equation}
i.e., $\dot{M}_{\rm acc}$ asymptotes to a $t^{-5/3}$ power law \citep{Michel1988,Chevalier1989b}. A simple 
injection model that exhibits this $t^{-5/3}$ asymptotic behaviour would have $k = 5/3$, and
\begin{align}
    \Eeng &\approx \eta c^2 \cdot \frac{2 (3 - m)}{3} M_{\rm enc}(r_{\rm acc}) \nonumber \\&\approx 
    2.4\e{53}~\left(\frac{\eta}{0.1}\right)\left(\frac{3-m}{2}\right)\left(\frac{M_{\rm 
    enc}}{M_\odot}\right)\unit{erg},
\end{align}
i.e., $M_{\rm acc} = 2 (3 - m) / 3 \cdot \Menc(r_{\rm acc}) \sim \Menc(r_{\rm acc})$ in Equation 
\ref{eq:accretion_lum}. The efficiency $\eta$ is not known (and may vary with time), but commonly adopted values 
are $10^{-3}$ for energy injection via a disc wind and $0.1$ for a collimated jet \citep{Kasen2017}.

\subsection{Dimensionless Parameters}\label{sec:dim_par}

We can use the ejecta parameters ($\Mej$, $\Ekej$) and engine parameters ($\teng$, $\Eeng$, $\Meng$)
to write down dimensionless quantities governing the evolution of the ejecta when there is energy input from
the central engine. Let us consider a unit system where the units of time, mass and energy are given by
\begin{equation}
    t_\sim = \teng,~~~~~~~~M_\sim = \Mej,~~~~~~~~E_\sim = \Ekej,
\end{equation}
and let $\tilde{Q}$ denote the magnitude of a quantity $Q$ in units where $t_\sim = M_\sim = E_\sim = 1$.
Our derived units for length, velocity, and density in this unit system are
\begin{align}
    r_\sim &= \teng \Mej^{-1/2} \Ekej^{1/2} \\
    v_\sim &= \Mej^{-1/2} \Ekej^{1/2}\\
    \rho_\sim &= \teng^{-3} 
    \Mej^{5/2} \Ekej^{-3/2},
\end{align}
where we note that $v_\sim = r_\sim / t_\sim = \zeta_E^{1/2} v_t$ (see Equation \ref{eq:E_to_M_ratio}).

If we neglect the effects of both gravity and radiation transport, the hydrodynamical evolution is mainly characterized by the ratio
\begin{equation}
    \Eengtild = \frac{\Eeng}{\Ekej},
\end{equation}
which sets the energy deposition rate. For the regimes considered in this paper, gravity is dynamically unimportant everywhere but the innermost regions of the ejecta near the central object, since the ejecta binding energy is small compared to its kinetic energy. Including the effect of radiative diffusion introduces an additional dimensionless parameter
\begin{equation}\label{eq:dimless_diff_time}
    \tilde{t}_d = \frac{t_d}{\teng} \sim \frac{\sqrt{\left(\kappa \Mej\right) / \left(v_t c\right)}}{\teng},
\end{equation}
where $t_d$ is the effective diffusion time in a homologously expanding medium \citep{Arnett1982}. For
$\tilde{t}_d \gg 1$, we expect radiative losses to be unimportant and a purely hydrodynamical description
to apply. Since our simulations in this study ignore radiation transport in the dynamical phase, they are
only strictly applicable in this regime. 

In these limits where gravity and radiation diffusion are unimportant, the character of the hydrodynamics
is determined solely by the one dimensionless parameter 
$\Eengtild = \Eeng/\Ekej$. Therefore we need only run simulation for a given value of $\Eengtild$ and can scale the result to specific values of $\Mej, \teng, \Ekej, \Eeng$ by a simple change of units.  

\subsection{Self-Similar Solution for Shock Evolution}\label{sec:analytic_sol}

For a spherically symmetric system, the initial dynamics will be described by self-similar analytics.  We expect that the central engine will inflate a bubble or cavity in the centre of the ejecta, and drive a shock into the outer regions \citep{Ostriker1971,Chevalier1977,Chevalier1992,Kasen2016}. Assuming that the interior of the engine-inflated bubble can be treated as a uniform fluid with adiabatic index $\gamma = 4/3$, that radiative losses are unimportant, and that the gas is swept into a thin shell, the evolution of the bubble can be described by thin shell momentum and energy equations
\begin{align}
    M_{\rm sh}\frac{dv_{\rm sh}}{dt} &= 4\pi R_{\rm sh}^2 \left[P - \rho_e (v_{\rm sh} - v_e)^2\right]\\
    \frac{d(4\pi R_{\rm sh}^3 P)}{dt} &= L - 4\pi R_{\rm sh}^2 P \frac{dR_{\rm sh}}{dt}.
\end{align}
Here, $M_{\rm sh}$ is the mass of the shell, $R_{\rm sh}$ is the position, $v_{\rm sh}$ is the velocity, $P$
is the internal pressure, $\rho_e$ is the external density, $v_e$ is the external velocity, and $L$ is the
engine luminosity.

We let $L = L_0 \left(\frac{t}{\teng}\right)^{-l} = \frac{\Eeng}{\teng} (k - 1) 
\left(\frac{t}{\teng}\right)^{-l}$, where $L_0$ is the initial luminosity, $l > 0$ is an index governing the luminosity 
decay, and $k = L_0 \times \teng / \Eeng + 1$. This approximates energy injection modelled by Equation \ref{eq:lum_eng} 
in the limits $t \ll \teng$ (where the injection rate is roughly constant, and $l = 0$) and $t \gg \teng$ ($l = k$), 
assuming that the engine turns on at $t = 0$. With this form for the luminosity, the thin shell equations admit a 
self-similar solution for $R_{\rm sh}$. We give the solution in the unit system defined in Section \ref{sec:dim_par}, 
which simplifies the formulae.

Inside the inner ejecta, where $\rho(R) \propto R^{-d}$, the solution for $\tilde{R}_{\rm sh}$ (valid for $l < 1$) takes the form
\begin{equation}\label{eq:R_sh_dimensionless}
    \tilde{R}_{\rm sh}(\ttild) = A\ttild^\alpha,
\end{equation}
where
\begin{equation}\label{eq:R_sh_coeff}
    A = \zeta_E^{-1/2} \left[\frac{\zeta_{\rm sh} \zeta_E}{\zeta_\rho} \Eengtild (k - 
    1)\right]^{1/[5-d]},
\end{equation}
the power-law index is
\begin{equation}\label{eq:R_sh_plaw_index}
    \alpha = \frac{6 - l - d}{5 - d},
\end{equation}
and we have defined
\begin{equation}\label{eq:zeta_sh_def}
    \zeta_{\rm sh} \equiv \frac{(5-d)^3(3-d)}{[(11 - 2d) - (6-d)l](1 - l)[(9 - 2d) - (4-d)l]}.
\end{equation}
For $l = 0$ and $\Rej \gg R_t$, this solution is equivalent to those of \citet{Chevalier1992} and \citet{Kasen2016} (see their equations 2.6 and 13 respectively). If there is minimal variation in the indices that set the energy deposition profile and the parameters controlling the structure of the ejecta, it is apparent from Equation \ref{eq:R_sh_coeff} that the shock evolution in this dimensionless space is set primarily by the energy ratio $\Eengtild$.

For $l \geq 1$, we expect the shell to coast at a constant velocity matching the background ejecta velocity 
\citep{Chevalier1992}. If $l < 1$ initially but $l \geq 1$ at late times, the shock would still expand 
superlinearly, with the upper bound on $R_{\rm sh}$ given by Equation \ref{eq:R_sh_dimensionless}. However, as $t \rightarrow \infty$, it would approach the constant velocity free expansion phase.

The time $\ttild_t$ required for the shock to reach the transition point in the velocity profile (again for $l < 1$) is 
\begin{equation}
    \ttild_t = \left[\frac{\zeta_\rho}{\zeta_{\rm sh} \zeta_E} \Eengtild^{-1} (k - 1)^{-1} \right]^{1/(1-l)}.
\end{equation}
If $\ttild_t \gtrsim 1$, the central engine energy ejection will wind down before the shock has a chance to 
reach the transition point, and the shell will enter free expansion while still embedded in the ejecta. However, if $\tilde{t}_t \lesssim  1$, we expect that the shock will reach the transition point and accelerate down the steep outer layers of ejecta, eventually breaking out of the star. This implies the condition
\begin{equation}
    \Eengtild \gtrsim \frac{\zeta_\rho}{\zeta_{\rm sh} \zeta_E} (k - 1)^{-1}
\end{equation}
for the shock to break out of the ejecta. For the parameters considered in this paper,
$\zeta_\rho/(\zeta_{\rm sh} \zeta_E)$ is order unity, so the shock breakout condition is roughly that the energy deposited by the central engine exceed the ejecta kinetic energy.

\section{Numerical Simulation Setup}\label{sec:simulation_setup}
    
We use the \castro\ hydrodynamics code \citep{Almgren2010,Almgren2020,Zingale2018} for  
our numerical calculations. Our simulation setup is publicly available on GitHub (see
the Data Availability section). The simulations are
performed in 2D axisymmetric coordinates $(r, z)$. We use outflow boundary conditions
except for the inner radial boundary, which is reflecting. While \castro\ stores all
components of the velocity regardless of dimensionality, here we disregard the azimuthal
component and set it to 0. The central engine is represented by a point gravitational mass
at the origin $(r = 0, z = 0)$ surrounded by an extended region where the engine provides
mass and energy source terms.

Below we describe our numerical approach and simulation setup. To distinguish the 
position vector from the radial coordinate, in the subsequent sections we will denote the 
position vector as $\pvec = (r, z)$ with magnitude $R = \sqrt{r^2 + z^2}$.

\subsection{Hydrodynamics}

\castro\ uses an unsplit version of the piecewise parabolic method \citep{Colella1984} for 
solving the hydrodynamics. It solves the compressible Euler equations with external source 
terms; we discuss the source terms introduced to simulate energy injection from the central
engine in Section \ref{sec:central_src}. We complete the system of equations with a gamma law
equation of state (EoS) for a monoatomic ideal gas, $P = (\gamma - 1) \rho e$.
Like \citet{Suzuki2017}, we assume that radiation pressure dominates and that gas and radiation 
are tightly coupled and so use $\gamma = 4/3$.

We determine the gravitational acceleration $\gvec$ using the monopole approximation, in
which the enclosed $\Menc$ is calculated by computing a 1D average of the mass density
and integrating it to find the acceleration, then interpolating the 1D acceleration profile
onto the simulation grid. We also incorporate the gravitational field from a point mass at
the origin representing the engine. As discussed in Section \ref{sec:dim_par}, gravity is
only dynamically important in the central region of the ejecta where $R \ll R_t$ and should
not impact the global simulation.

\subsection{Initial Conditions}

We initialize the density profile of the ejecta with the broken power-law profile
(Equation \ref{eq:init_dens_prof}) with power-law exponents $d = 1$ and $n = 10$.
In all simulations we choose an ejecta mass of $\Mej = 4\unit{M_\odot}$ and a
velocity and the transition point of $v_t = 0.02c$, yielding a kinetic energy of
$\Ekej \approx 10^{51}\unit{erg}$. These default values are characteristic of the
supernova explosion of a stripped-envelope star, but as discussed in Section
\ref{sec:dim_par}, our results can be rescaled to other values of the ejecta
properties by a change of units. We begin our simulations after the ejecta has
already expanded homologously for a time $t_0 = 0.1~\teng$. This start time is
early enough in the evolution that we do not expect the central engine to have
significantly impacted the ejecta, but late enough that the ejecta has grown
sufficiently large for us to spatially resolve its structure. We assume that the
kinetic energy of the ejecta dominates the thermal energy by the start of the
simulation (i.e.\ that any significant initial thermal energy has b een adiabatically
degraded). We thus simply use an initial temperature value of
$T_{\rm ej} = 1000\unit{K}$, which gives us $\Ethej \ll \Ekej$ as required.

The region of the simulation domain initially outside the ejecta ($R > \Rej$) is taken
to be a uniform density, stationary ambient medium. We choose the ambient density
$\rho_a$ such that the total mass of the ambient medium is only $1$ per cent of the ejecta
mass ($0.04\unit{M_\odot}$). 
This ensures that the ejecta will be minimally affected
by interaction with the ambient medium. The radius, $\Rej$, where the ejecta profile
transitions to the ambient medium is set such that $\rho(\Rej) = \rho_a$, which
gives $\Rej$ a factor of about ten times larger than the transition radius $R_t$ for
our setup. The temperature of the ambient medium is initialized to $T_a = 100\unit{K}$.

We advect several species along with the flow to investigate chemical mixing and serve
as approximate tracers for different fluid elements. From a hydrodynamic standpoint, it
is irrelevant which species we use for this purpose. One element with mass fraction
$X_{\rm core}$ comprises the inner $12.5$ per cent ($0.5\unit{M_\odot}$) of the ejecta, another
element with mass fraction $X_{\rm ims}$ the next $25$ per cent ($1\unit{M_\odot}$),
and the remaining $62.5$ per cent is the element assigned to the envelope (mass fraction $X_{\rm env}$;
$2.5\unit{M_\odot}$). The element shells are blended at their edges, i.e., there is an exponential
dropoff rather than an immediate cutoff in an element's mass fraction at the edge of it's shell.
The wind from the central engine (see Section \ref{sec:central_src})
is injected as a separate tracer element ($X_{\rm wind}$), which allows us to trace particles in
the wind. This distribution of elements is not intended to be a realistic representation of a
true supernova, but is used to study how layers of ejecta are mixed and elements
redistributed in the dynamics. 

\subsection{Central Energy Source}\label{sec:central_src} 

The central engine is simulated by including an energy source at the centre
of the domain, which takes the form of a volumetric luminosity source term
$\mathcal{L}(\pvec, t)$ in the energy equation.
The size of the deposition region is initially roughly $3$ per cent of the
ejecta radius, and increases with time at a rate of $0.03~v_t$ such that the deposition
region scales proportionally with the initial expansion of the ejecta. The energy is
injected purely as thermal energy throughout the deposition region, with a smooth
exponential cutoff deposition rate at the edge. The resulting pressure gradient drives
a wind that flows outward from the deposition region, adiabatically converting the
thermal energy into kinetic energy. 

The engine luminosity, $\Leng$, is that of Equation \ref{eq:lum_eng} with $t = 0$
corresponding to the start of the simulation. To prevent small time steps and
excessive subcycling in time due to rapid evacuation of the deposition region, we
begin the simulation with a linear ramp-up of $\Leng$ over a time-scale $t_r = 0.05~\teng$.
Since $t_r \ll \teng$ the ramp-up only leads to a $\sim 2.4$ per cent reduction in the
total energy emitted by the engine. 

For most of our simulations, we inject energy from the central engine symmetrically about
the origin (an isotropic energy injection). The volumetric deposition rate in a given cell
within the deposition region then obeys $\mathcal{L}_{\rm iso} \propto \Leng V_{\rm dep}^{-1}$,
where $V_{\rm dep}$ is the volume of the deposition region. However, wind from a physical engine
would likely exhibit some anisotropy. To explore the impact of this anisotropy, we consider
several other deposition schemes. In one instance, the volumetric deposition rate is
$\mathcal{L}_{\sin^2} \propto \mathcal{L}_{\rm iso} \sin^2 \theta$, where $\theta$ is the
polar angle corresponding to the cell. This is consistent with magnetar dipole spin-down
where the dipole moment is directed along the vertical axis. We also consider the case of
purely equatorial energy injection, where all energy from the engine is deposited within
$\pm 5 \degree$ of the radial axis. Finally,
we consider a case with deposition rate $\mathcal{L}_{\cos^2} \propto \mathcal{L}_{\rm iso}
\cos^2 \theta$, where energy is preferentially deposited in the polar direction. This may
better approximate injection via weak jets from a central magnetar or black hole.

Like \citet{Chen2016,Chen2017,Chen2020}, we add mass to the deposition region to prevent it 
from evacuating completely and severely limiting the timestep. For our models the mass
injection rate is proportional to the energy injection rate, and is taken to be
\begin{equation}\label{eq:mass_inj}
    \dot{\rho} = \mu \frac{\mathcal{L}}{c^2},
\end{equation}
where $\mu$ is a dimensionless parameter controlling the injection rate and $c$ is the speed
of  light in a vacuum. The injected mass is assumed to be cold and does not introduce additional
thermal energy. To avoid the clumping up of the injected material in the innermost cells at
early times and instead produce a steady wind, we add the mass at the escape velocity of the
central object evaluated in the innermost simulation cell. This is implemented as a momentum
and kinetic energy source, and the added energy is negligible compared to the engine energy
and ejecta kinetic energy.

We set $\mu = 7.5$ for most of our simulations, as this prevents both the wind velocity and 
soundspeed from becoming superluminal while keeping the amount of injected mass under $\sim 
2$ per cent of the ejecta mass. For the highest energy cases we choose a smaller $\mu$ and permit
the velocity to go superluminal ($\mu \approx 2.08$ for the \textsf{7/16ms} run, and
$\mu \approx 1.70$ for the \textsf{1/4ms} run), since the amount of mass injected would
otherwise approach $10$ per cent of the ejecta mass and may begin to impact the supernova dynamics.

\subsection{Simulation Grid}\label{sec:grid}

We  use  adaptive mesh refinement (AMR; see \citealt{Berger1989}) to reduce the 
computational cost of the simulations. All of our simulations have a base grid of size $256 
\times 512$, with $3$ levels of refinement that each increase the resolution by a factor of 
$4$. The peak resolution is thus equivalent to $16384 \times 32768$. The spatial resolutions
$\Delta R$ range from $9.1\e{8}\unit{cm}$ to $6.1\e{11}\unit{cm}$; the ratio $\Delta R/R_t$
falls in the range $\numrange{0.037}{0.055}$ for all runs.

The ejecta begin at one level of refinement above the base grid, and the refinement region
expands with time to guarantee that it always encapsulates the ejecta. We set the scale of the
highest-resolution AMR level so that it contains the energy deposition region. We also let it
expand with time following Equation \ref{eq:R_sh_dimensionless}, which ensures that any central
cavity that develops will be placed at the highest possible resolution. Finally, we use an
additional flag based on composition to maximally resolve any regions containing elements from
the inner ejecta or engine-generated wind. This also enables us to capture any wind breakout
into the outer ejecta at high resolution.

\section{Results of Hydrodynamics Simulations}\label{sec:results}
    
We ran a suite of 10 simulations, covering a range of engine parameters. We provide a
summary of simulation parameters and simulation names/IDs for referencing them in text
in Table \ref{tab:sim_par}. Nine of the models adopt the energy injection formula of
magnetar dipole spin-down. For six of these runs (those whose IDs simply contain the
rotation period), we assume an effective external dipole field strength of $10^{15}\unit{G}$
and inject energy isotropically, varying only the magnetar rotation period.

\markchange{Through the rescaling discussed in Section \ref{sec:dim_par}, these models can be
used to study a wider range of model parameters. In particular, the \textsf{1/4ms} and
\textsf{7/16ms} runs can be used to study scenarios with engine-to-ejecta energy ratios
$\Eengtild \gtrsim 100$, even though the nominal spin periods may be unphysical for a neutron
star (i.e., above the mass-shedding limit). These include scenarios with extreme energy reservoirs
$\Eeng \gtrsim 10^{53}\unit{erg}$ due to large magnetar masses (see Section 4 of
\citealt{Metzger2015}), and scenarios with low ejecta kinetic energies and masses.
Models with low ejecta masses can be difficult to directly simulate, as the
mass injected to limit the wind velocity can become a significant fraction of the ejecta
mass and impact the dynamics.}

We examine the effect of lowering the magnetic field strength (increasing $\teng$) in
our \textsf{1ms\_lowB} model, which corresponds to a magnetar with a $1\unit{ms}$
rotation period and dipole field strength of $2.5 \times 10^{14}\unit{G}$. We also
introduce  asymmetry in the energy injection scheme in 3 of our models
(\textsf{1ms\_sin}, \textsf{1ms\_eq}, \textsf{polar\_5/3}; see Section \ref{sec:central_src}).
The last of these is intended to simulate, very roughly, a black
hole-like engine; it has the same engine time-scale and energy reservoir as our
\textsf{1ms} magnetar model, but incorporates a slower luminosity decay more
consistent with fallback accretion, preferentially deposits energy in the polar
direction, and has a larger central point mass.

The results of the \textsf{1/4ms}, \textsf{7/16ms}, and \textsf{2ms} runs
do not lead to any additional conclusions on top of those presented here,
and are not discussed. The evolution of the \textsf{1ms\_lowB} run strongly
resembles the \textsf{1ms} run (see the rescaling discussion in
Section \ref{sec:dim_par}), so it is only discussed in Section
\ref{sec:observables}. However, data and plots from these runs will be shared
on reasonable request to the corresponding author.

\begin{table*}
    \caption{Summary of simulation runs with selected parameters for each. The parameters are the engine 
            time-scale $\teng$, the engine to ejecta energy ratio $\Eeng/\Ekej = \Eengtild$, the end time of the 
            simulation $t_{\rm max}$, the maximum extent of the simulation domain $R_{\rm max}$, the luminosity decay index $k$, the mass ratio between the central point mass and the ejecta $\Meng/\Mej = \tilde{M}_{\rm eng}$, and the energy deposition scheme (see Section \ref{sec:central_src}).\label{tab:sim_par}}
    \begin{tabular}{cccccccc}
    \hline\hline
    Simulation ID & $\teng\left(\mathrm{h}\right)$ & $\Eeng/\Ekej$ & 
        $t_{\mathrm{max}}\left(\teng\right)$ & $R_{\mathrm{max}}\left(\mathrm{cm}\right)$ & 
        $k$ & $\Meng/\Mej$ & Deposition Scheme\\
    \hline
    \textsf{1/4ms} & $0.036$ & $315.2$ & 2.0 & $5.00 \times 10^{12}$ & $2$ & $0.3$ & Isotropic\\
    \textsf{7/16ms} & $0.109$ & $102.9$ & 2.0 & $1.50 \times 10^{13}$ & $2$ & $0.3$ & Isotropic\\
    \textsf{1ms} & $0.569$ & $19.7$ & 5.0 & $7.50 \times 10^{13}$ & $2$ & $0.3$ & Isotropic\\
    \textsf{2ms} & $2.275$ & $4.9$ & 12.0 & $3.75 \times 10^{14}$ & $2$ & $0.3$ & Isotropic\\
    \textsf{3ms} & $5.119$ & $2.2$ & 21.0 & $10^{15}$ & $2$ & $0.3$ & Isotropic\\
    \textsf{10ms} & $56.875$ & $0.2$ & 33.0 & $10^{16}$ & $2$ & $0.3$ & Isotropic\\
    \textsf{1ms\_lowB} & $9.100$ & $19.7$ & 5.0 & $1.20 \times 10^{16}$ & $2$ & $0.3$ & Isotropic\\
    \textsf{1ms\_sin} & $0.569$ & $19.7$ & 5.0 & $7.50 \times 10^{13}$ & $2$ & $0.3$ & $\mathcal{L} \propto \sin^2 
    \theta$\\
    \textsf{1ms\_eq} & $0.569$ & $19.7$ & 7.5 & $7.50 \times 10^{13}$$^{^*}$ & $2$ & $0.3$ & Equatorial\\
    \textsf{polar\_5/3} & $0.569$ & $19.7$ & 5.0 & $7.50 \times 10^{13}$ & $5/3$ & $1.0$ & $\mathcal{L} \propto 
    \cos^2 
    \theta$\\
    \hline\hline
    \end{tabular}
    \vspace{0.5cm}\\
    $^*$ Vertical extent in each direction; the simulation domain for this run is elongated in the horizontal direction to $1.125\e{14}\unit{cm}$.
\end{table*}

\subsection{Evolution of the Fiducial Simulation}

Our \textsf{1ms} model represents a case where the central engine dominates
the energetics, but the energy injection rate is not so high that we need a large
injected mass ($\gtrsim 0.02~\Mej$) to prevent superluminal wind velocities. We take
this to be our fiducial model, and explore the dynamics in this section.

\subsubsection{Development and Evolution of Central Bubble}

\begin{figure}
    \centering
    \plotone{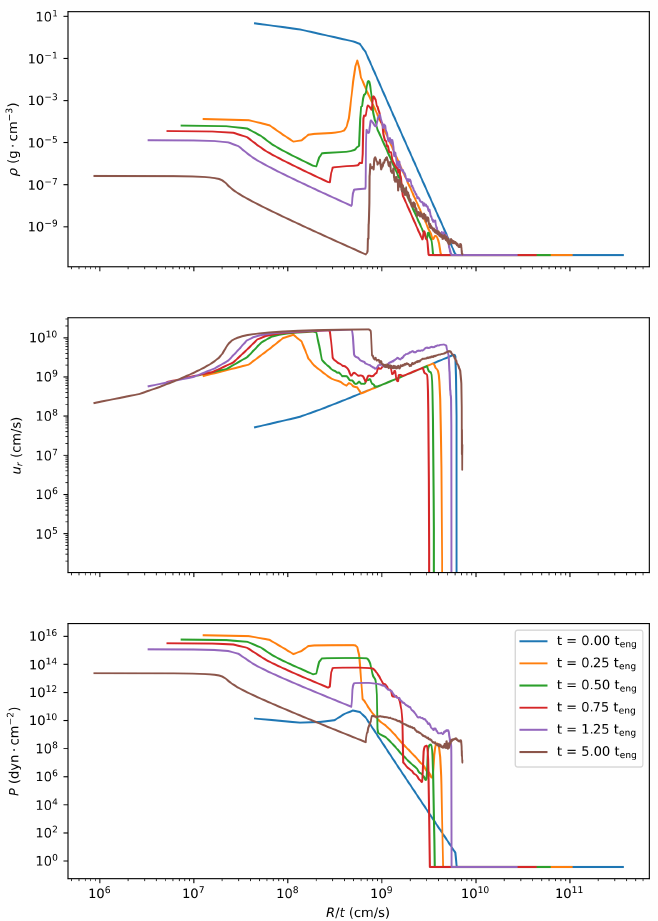}
    \caption{1D angle-averaged density, radial velocity, and pressure profiles (from top to bottom) for our 
    \textsf{1ms} simulation run, plotted at different time points.}
    \label{fig:profile_plots}
\end{figure}

\begin{figure*}
    \centering
    \plotone{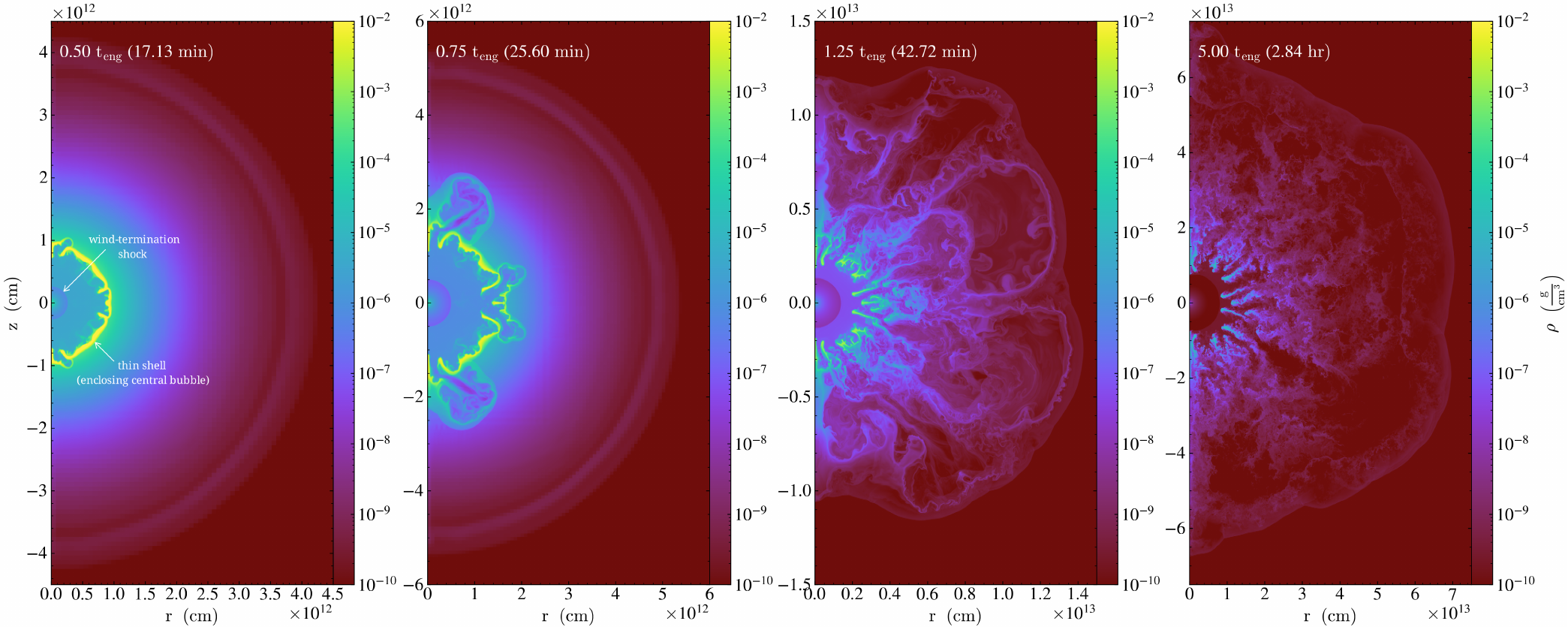}
    \caption{Density maps showing the time-evolution of the \textsf{1ms} simulation. The energy injection from the
    central engine inflates a bubble in the centre of the ejecta, filled with high-pressure gas. Rayleigh-Taylor
    instabilities develop at the edge of the bubble, eventually causing it to rupture at discrete points along
    its surface. The gas contained in the bubble vents out through these rupture points and leaves behind
    low-density channels in the remnant.}
    \label{fig:1ms_dyn}
\end{figure*}

\begin{figure}
    \centering
    \plotone{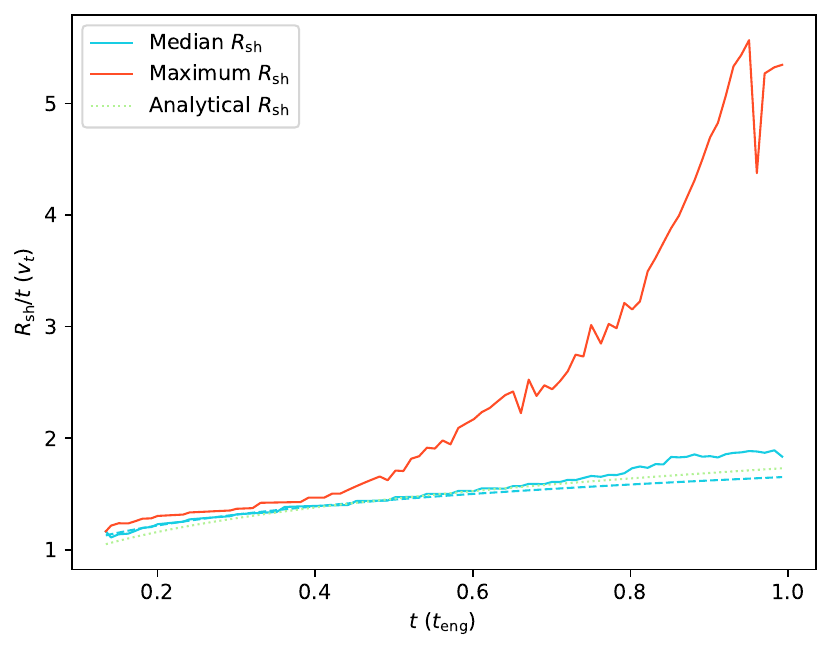}
    \caption{Shock position (in velocity space) vs.\ time for our \textsf{1ms} model. The solid lines show the 
    median shock position and maximum shock position across all angles; the difference between the two is a result 
    of the growth of instabilities and the eventual breakout of the gas from the central cavity (the top line
    corresponding to the maximum shock position tracks the initial breakout). The dashed line shows a power-law
    fit to the shock position using data from $t < 0.5~\teng$. The dotted line is the analytical prediction for
    a 1D model given by Equation \ref{eq:R_sh_dimensionless}.}
    \label{fig:shock_pos_vs_time_example}
\end{figure}

The evolution of our fiducial \textsf{1ms} model is shown in Figure \ref{fig:profile_plots}
and Figure \ref{fig:1ms_dyn}. As seen in the angle-averaged density profiles in Figure \ref{fig:profile_plots}, the
wind from the central engine excavates a low-density cavity or bubble in the centre of the
ejecta, sweeping the gas into a thin shell at the edge of the cavity. In 1D simulations
(e.g., \citealt{Kasen2010}), the shell is stable and is driven outward in a spherically-symmetric
manner by the high pressure inside of the cavity. In 2D, the low-density wind colliding with the
higher density material at the edge of the bubble produces Rayleigh-Taylor (RT) instabilities behind
the shock front. These instabilities cause the surface of the shell to flex and RT fingers to develop,
consistent with previous 2D studies (see, e.g.\ \citealt{Chen2016,Blondin2017,Suzuki2017,Suzuki2021b}).
We rely on numerical diffusion to seed the instabilities, which may delay the growth somewhat compared
to initial conditions with small-scale density variation. The interior of the cavity is not entirely
homogeneous as was assumed in Section \ref{sec:analytic_sol}: a ``wind-termination shock" develops where
the high-velocity wind from the deposition region meets the gas in the outer regions of the cavity. The shell
and wind termination shock are labelled in the first plot in Figure \ref{fig:1ms_dyn}.

In the phases while the shell remains in the inner ejecta ($R_{\rm sh} < R_t$), we expect
its expansion to initially follow the analytic solution (Equation \ref{eq:R_sh_dimensionless}),
and then begin to tend toward free expansion as $t$ becomes comparable to $\teng$. Once
the shell begins to enter the coasting/free expansion phase, the analytic solution gives an
upper limit on the bubble radius. However, if the shell has crossed into the steep outer layers
of the ejecta ($R_{\rm sh} \geq v_t t$), the runaway growth of instabilities and resultant
asphericity in the shell eventually render a 1D inadequate to describe the shell location. We
plot the shell/bubble radius as a function of time in Figure \ref{fig:shock_pos_vs_time_example}.
We measure the shock position numerically as a function of angle by applying the ridge/vessel
detection filter of \citet{Frangi1998} as implemented by \textsc{scikit-image}
\citep{vanderWalt2014} to density data from the simulation. We see that the median shock position
across all angles is reasonably well-approximated by the analytic solution out to $t \sim 0.8~\teng$,
but the maximum shock position quickly begins to deviate from the analytic expectation due to
the growth of instabilities. Due to the decay in engine luminosity, a power-law fit to the median
shock position up to $t = 0.5~\teng$ returns an index $\alpha \sim 1.19$, slightly less than the
$\alpha = 1.25$ predicted by Equation \ref{eq:R_sh_plaw_index} for a constant luminosity source.
We also note that the finite size of the energy deposition region might cause the shell to
initially form at a larger radius than expected, which would contribute to the discrepancy
between the measured shock position and the analytic solution at early times.

\subsubsection{Breakout from the Central Bubble}\label{sec:bubble_breakout}

Under pressure from the gas inside the cavity, the shell is driven into the outer region of the
ejecta, where the expansion accelerates due to the steeper density gradient. Because of the RT
instabilities, this transition occurs non-uniformly across the shell. This causes ``lobes" or
``fingers" containing high-pressure, low-density gas emerge from the shell and expand in the
outer ejecta. The bubble then ruptures, allowing the gas to vent out through the gaps in the
shell and disrupt the entire ejecta. This process is illustrated in the first three panels in
Figure \ref{fig:1ms_dyn}.

This breakout of the gas from the central cavity occurs asynchronously. In our \textsf{1ms} run,
gas bubbles first erupt from the shock front at roughly $10\degree$ from the vertical axis at a
time $t = \numrange{0.45}{0.60}~\teng$; this is followed by an equatorial breakout at $t = 0.65-0.80~\teng$.
The gas that rushes out through these channels is able to disrupt the structure of the entire outer
ejecta by $t = 1~\teng \approx 34\unit{min}$. Several additional breakouts occur at $\sim 1~\teng$
and between $1$ and $2~\teng$ and further fragment the shell, leaving behind only the high-density 
edges of the outflow channels. At the end of the simulation (at $t = 5~\teng$ in this instance), there
is still some energy injection and venting of gas, but most of the engine's energy has already been
deposited and the ejecta has settled into near-homologous expansion.

As the initial ejecta is spherically symmetric and the energy deposition is isotropic, the geometric
pattern of instabilities and outflows is likely a consequence of the underlying grid structure and resultant
numerical error. We expect this breakout pattern to show some differences if the simulation is performed in
3D due to the different turbulent cascade (see \citealt{Blondin2017,Suzuki2019,Chen2020}). Including additional physics in the simulation, e.g., radiation and magnetic fields, may also alter the breakout pattern.

\begin{figure}
    \centering
    \plotone{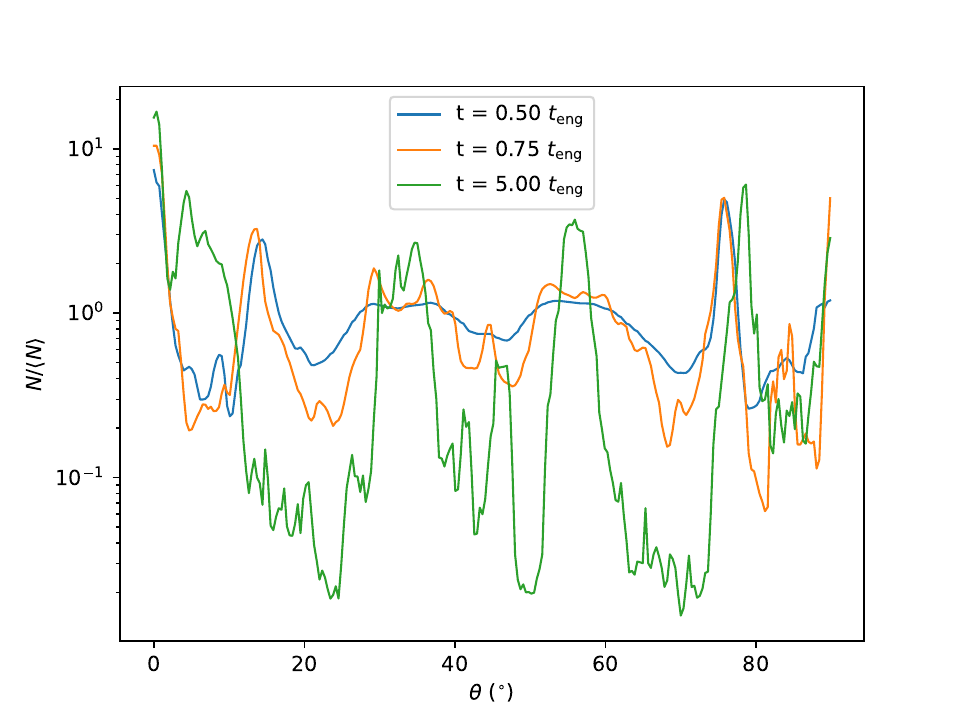}
    \caption{Angular variation in estimated column density (assuming pure hydrogen composition) for the 
    \textsf{1ms} simulation, rescaled for each time point to ease comparison. The column density is calculated 
    along rays of length equal to the radial extent of the domain. As the breakout is approximately symmetric, we only show variation over the top half of the domain. We can see that strong variation in column density (by several orders of magnitude) develops due to gas breakout out of the central cavity and forming channels in the outer ejecta.}
    \label{fig:N_vs_theta}
\end{figure}

One significant consequence of the breakout is the formation of relatively low-density
channels in the outer ejecta. To illustrate this, we plot column density as a function
of angle in Figure \ref{fig:N_vs_theta}. We find that dips in column density appear at
the angles where the gas breaks through the shell, with the variation eventually
spanning several orders of magnitude. Similar blowout-induced column density variation
was observed in \citealt{Blondin2017}. They also compared their 2D results to the results
of their 3D simulations, and found that the overall blowout structure is similar in 3D,
but small-scale features fill in the channels somewhat due to a larger contribution
from higher order instability modes. These channels could provide an avenue for radiation
from the engine to leak out of the ejecta and into interstellar space. 

\begin{figure}
    \centering
    \plotone{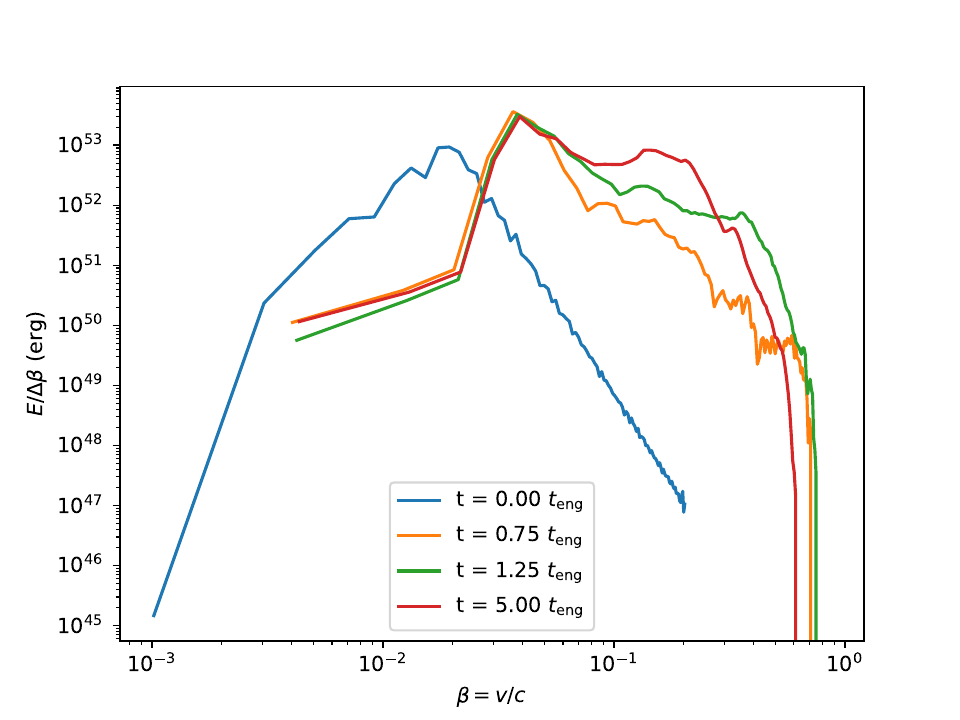}
    \caption{Distribution of total energy with velocity for our \textsf{1ms} model at 4 different time points 
    (including the initial and final step). This demonstrates the acceleration of the outer ejecta due to the gas 
    venting out of the central bubble. The interior of the central cavity is excluded when calculating the 
    distribution.}
    \label{fig:energy_vs_velocity}
\end{figure}

The breakout of the wind from the cavity also substantially accelerates the outer ejecta.
Figure \ref{fig:energy_vs_velocity} shows the distribution of energy in the ejecta
as a function of velocity, excluding the interior of the cavity. Initially, the bulk of the
energy is in material moving at velocities $\sim 0.02c$. Following the breakout, the energy
is largely distributed over material moving at velocities $\sim \numrange{0.05}{0.2}c$, with peak velocities
$> 0.5c$. This acceleration will decrease the effective photon diffusion time through the ejecta
(see Equation \ref{eq:dimless_diff_time}), and thus reduce the rise time of the light curve.
The observational consequences are discussed in more detail in Section \ref{sec:observables}.

The amount of energy in material moving at relativistic or mildly relativistic speeds ($v \gtrsim 0.3c$)
is difficult to quantitatively estimate for several reasons. The highest velocity ejecta in the
simulation experiences a deceleration due to interaction with the finite density circumstellar medium (CSM).
This deceleration is more extreme than it might be in, e.g., a lower density CSM or a ``wind-like" density profile with $\rho \propto R^{-2}$. We also do not include special relativity in our simulations, and the mass added to the grid to prevent superluminal velocities may cause us to overestimate the density of the engine wind. With these caveats, we can estimate that $\sim 10^{51}\unit{erg} \approx 0.05~\Eeng$ of energy is present at velocities $v \gtrsim 0.3c$ at $t = 1.25~\teng$ (shortly after the outflows catch up to the outer shock).

If accounted for, radiative diffusion could sap energy from the gas in the cavity
and outflows, leading to lower kinetic energy of the outflow. However, for the regime considered here
where $\tilde{t}_d \gg 1$ and $\Eengtild \gg 1$,
the breakout happens early enough that radiative losses should be minimal and the ejecta at the interface
with the CSM can plausibly reach relativistic speeds. This has implications for the the radio signatures of
engine-powered supernovae, as the subsequent interaction of the high velocity gas with the CSM is likely
to lead to radio synchrotron emission \markchange{(see also the discussions of radio signatures in 
\citealt{Suzuki2018,Suzuki2019,Suzuki2021b}, who performed simulations with special relativity)}.

\subsubsection{Element Mixing}

\begin{figure*}
    \centering
    \plottwo{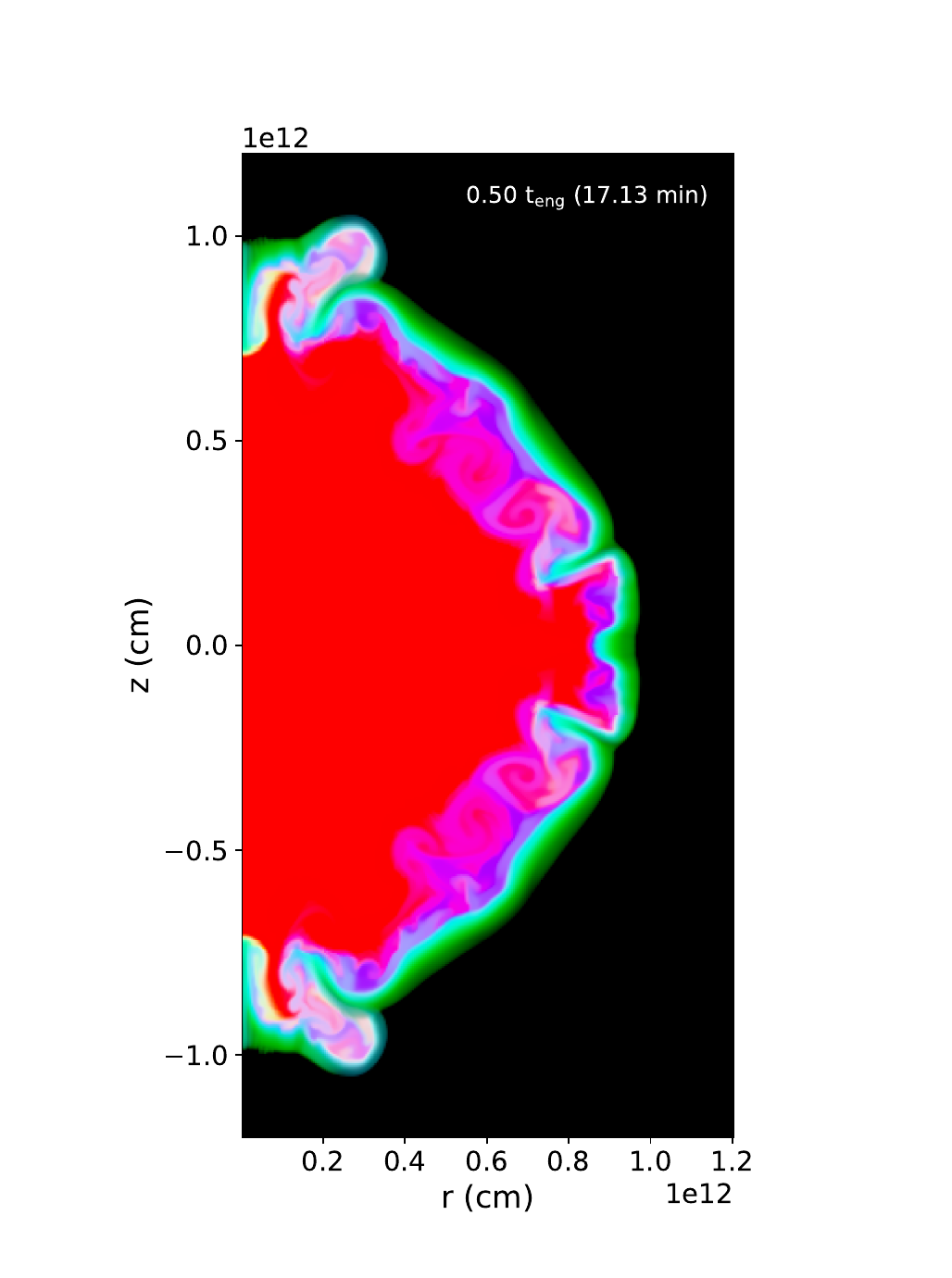}{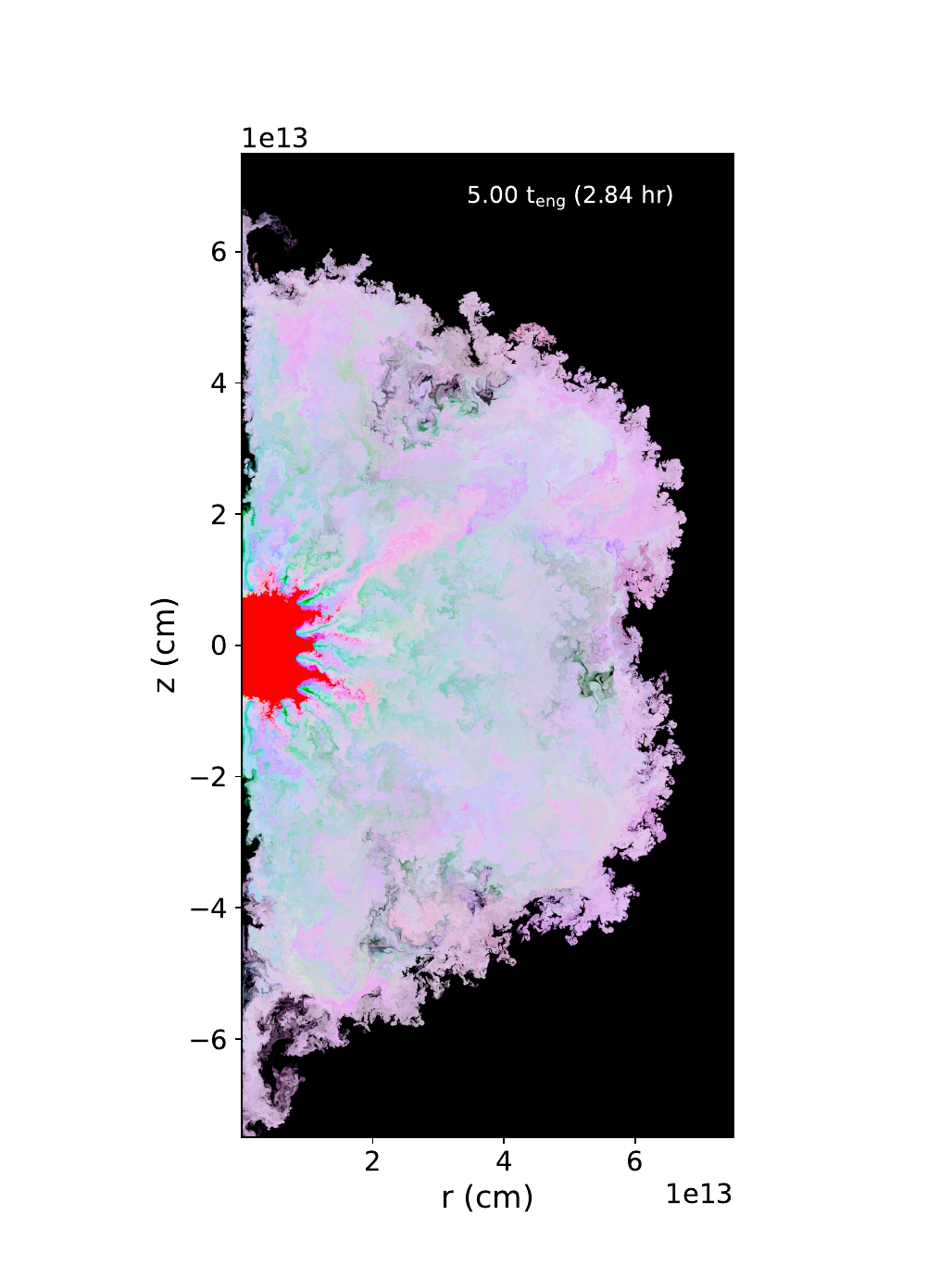}
    \caption{Composition split into RGB channels at $t = 0.5~\teng$ (left) and $t = 5.0~\teng$ (right) for our 
    \textsf{1ms} model. We map the mass fraction of the wind tracer element ($X_{\rm wind}$) to red, intermediate
    shell element ($X_{\rm ims}$) to green, and core element ($X_{\rm core}$) to blue, 
    and plot the mass fractions on a logarithmic scale from $10^{-5}$ to $1$.}
    \label{fig:composition_1ms}
\end{figure*}

\begin{figure*}
    \centering
    \plottwo{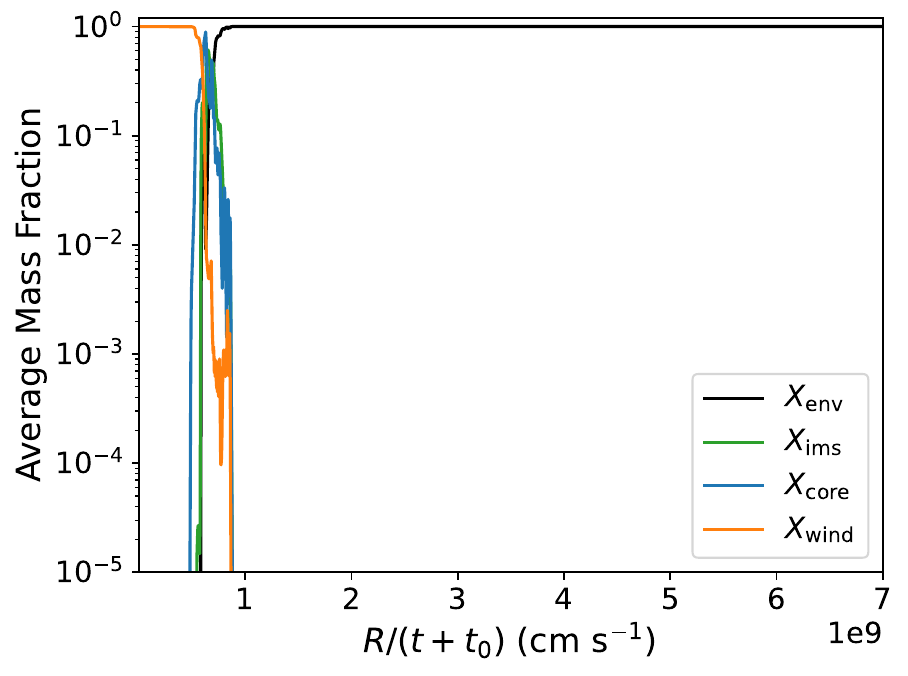}{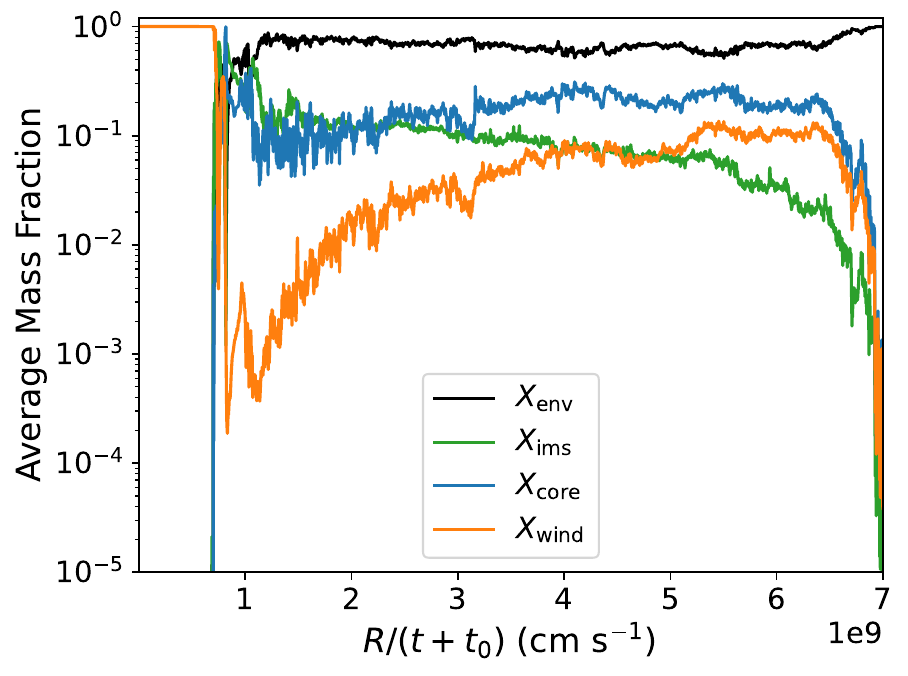}
    \caption{Angle-averaged distribution of species in the ejecta at $t = 0.5~\teng$ (left) and $t = 
        5~\teng$ (right) for our \textsf{1ms} model. The average is mass-weighted. 
        The x-axis gives the corresponding velocity coordinate, which is the radial distance 
        from the origin $R$ divided by the time elapsed. Elements swept up into the shell
        by the engine activity are blown out into the outer ejecta when the shell disintegrates
        and the wind from the engine breaks out.}
    \label{fig:comp_prof_1ms}
\end{figure*}

We also investigate the mixing of elements in the ejecta due to the additional
dynamics introduced by the central engine. The mixing process occurs in two main
stages. In the left panel of Figure \ref{fig:composition_1ms}, we
plot the distribution of elements at $t = 0.5~\teng$ (around the time of the first
breakout), zooming in on the central cavity. The region interior to the cavity
consists entirely of particles from the wind produced by the central
engine ($X_{\rm wind}$ is mapped to the red channel). The elements in the inner
ejecta are swept up into the shell. The initial onion-like structure is distorted
and mixed, but is still somewhat visible in the composition of the bubble before it
ruptures: the tracer element from the core of the original ejecta (coloured blue) is
concentrated at the inner edge of the shell
and in the RT plumes, while the tracer element for the intermediate zone between
the core and the envelope (green) is concentrated near the outer edge of the shell.

During the breakout from the central cavity, the elements in the shell are swept up with
the outflowing gas and dispersed throughout the outer ejecta. This mixing can be seen in
the map of the final distribution of elements shown in Figure \ref{fig:composition_1ms},
and the 1D mass fraction profiles in Figure \ref{fig:comp_prof_1ms}. Similar mixing
(described as a ``chemical inversion") was observed in \citet{Suzuki2021b}. We note that
$X_{\rm core}$ is higher than $X_{\rm ims}$ in the outermost regions of the ejecta,
despite the fact that the element corresponding to $X_{\rm core}$ initially comprised
the innermost layer. This could be a result of the stratification of elements swept
into the shell, as outflows will sweep up material from the inner edge of the shell
(where $X_{\rm core} > X_{\rm ims}$) and transport it to the region just behind the outer
shock front.

\subsection{Effect of Varying Energy Ratio}

\begin{figure*}
    \centering
    \plotone{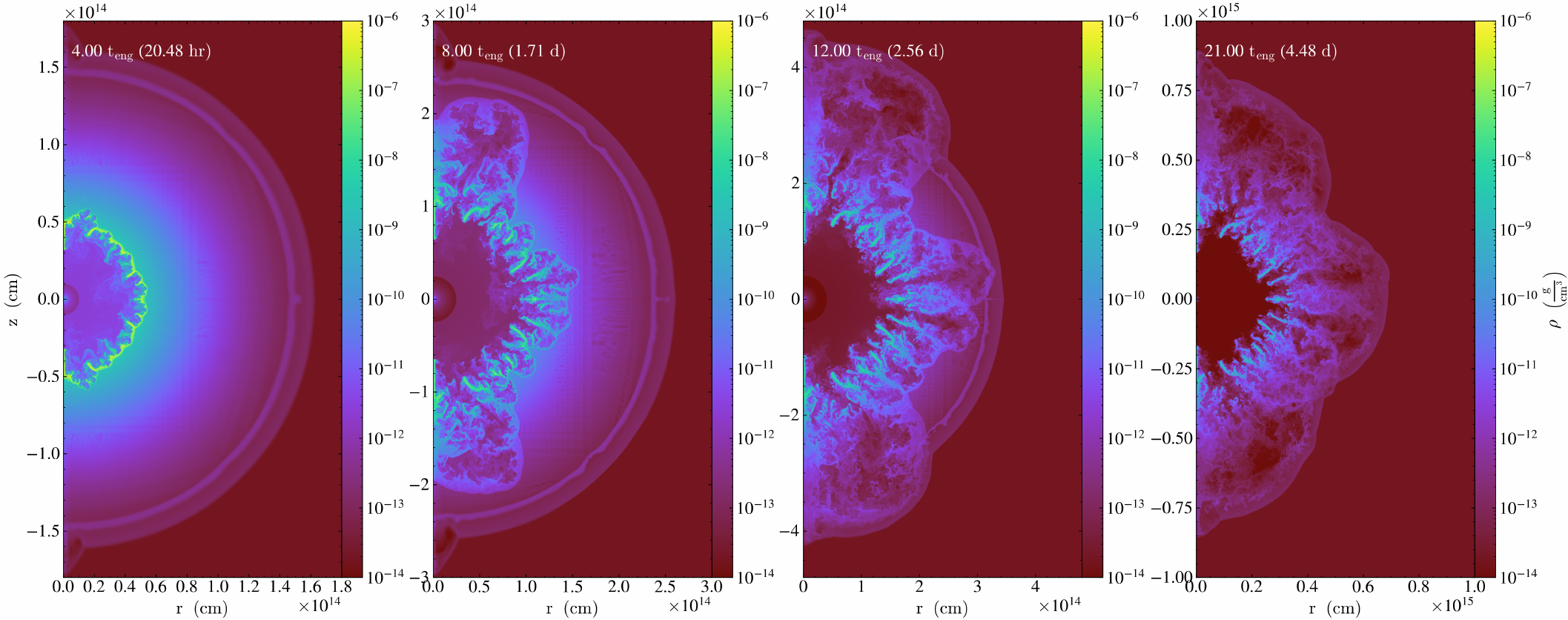}
    \caption{Density maps showing the time-evolution of the \textsf{3ms} simulation.}
    \label{fig:3ms_dyn}
\end{figure*}

\begin{figure}
    \centering
    \plotone{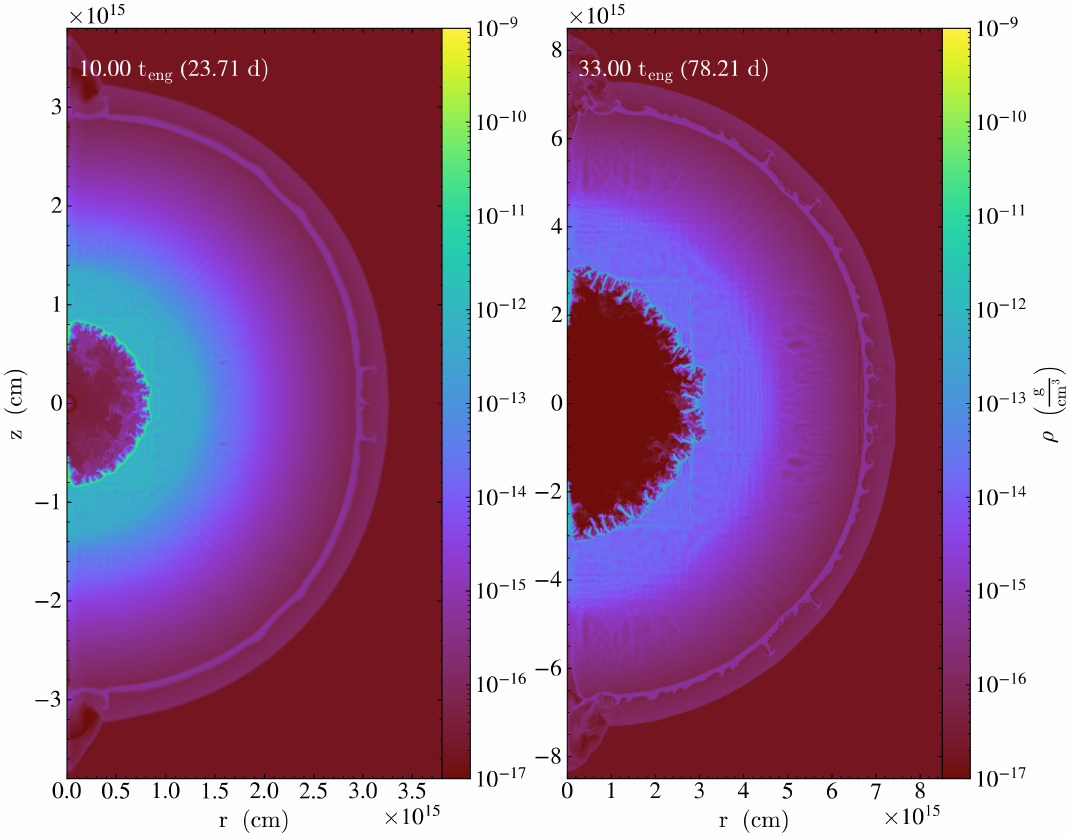}
    \caption{Density maps showing snapshots of the \textsf{10ms} simulation at $t = 10~\teng$ and at the
        final timestep ($t = 33~\teng$).}
    \label{fig:10ms_dyn}
\end{figure}

The primary dimensionless parameter governing the evolution of the system is the energy ratio
$\Eengtild = \Eeng/\Ekej$. To explore the impact of varying this parameter, we ran simulations
with energy ratios $\Eengtild$ ranging from $\approx \numrange{0.2}{315.2}$. In this section, we
examine two additional models -- the \textsf{3ms} simulation with
$\Eengtild \approx 2.2$, and the \textsf{10ms} simulation with $\Ekejtild \approx 0.2$ -- and
compare these to the \textsf{1ms} model, highlighting key differences in their evolution.

The hydrodynamical evolution of the \textsf{3ms} simulation is shown in Figure \ref{fig:3ms_dyn}.
Due to the smaller energy ratio, the breakout of the gas from the central cavity is less
catastrophic than in the \textsf{1ms} case. An initial breakout from the bubble occurs at $t =
\numrange{3.5}{4.5}~\teng$ and a second at $t =\numrange{5.5}{6.5}~\teng$, at similar polar 
angles to the \textsf{1ms} run breakouts. At $t \approx 9~\teng \approx 1.9\unit{d}$, the first 
plume breaks through the outer shock, and by $17~\teng \approx 3.6\unit{d}$ the outflows have 
consumed the entire outer ejecta. The breakout pattern leaves a visible imprint on the final
structure (the lobe pattern seen in the final panel of Figure \ref{fig:3ms_dyn}).

As seen in Figure \ref{fig:10ms_dyn}, the central engine in the \textsf{10ms} simulation is too 
weak to fracture the shell and drive outflows into the outer ejecta. Although RT fingers still
arise, the ejecta settle into homologous expansion with the swept-up shell still intact. The final
bubble radius is only $\approx 45$ per cent of the total ejecta radius.

In general, we expect that models with $\Eengtild \gtrsim 1$ will achieve breakout of the wind 
from the ejecta, and that models with $\Eengtild \lesssim 1$ will not. For $\Eengtild \lesssim 1$, the shell remains unshredded and we do not see bubble breakouts, low-density channels, or substantial acceleration of the outer ejecta. Asymmetry in the energy injection from the engine, especially if it comes in the form of a collimated outflow, could allow the engine to disrupt the outer ejecta even with a smaller energy budget.

\subsection{Anisotropic Deposition}

\begin{figure*}
    \centering
    \plotone{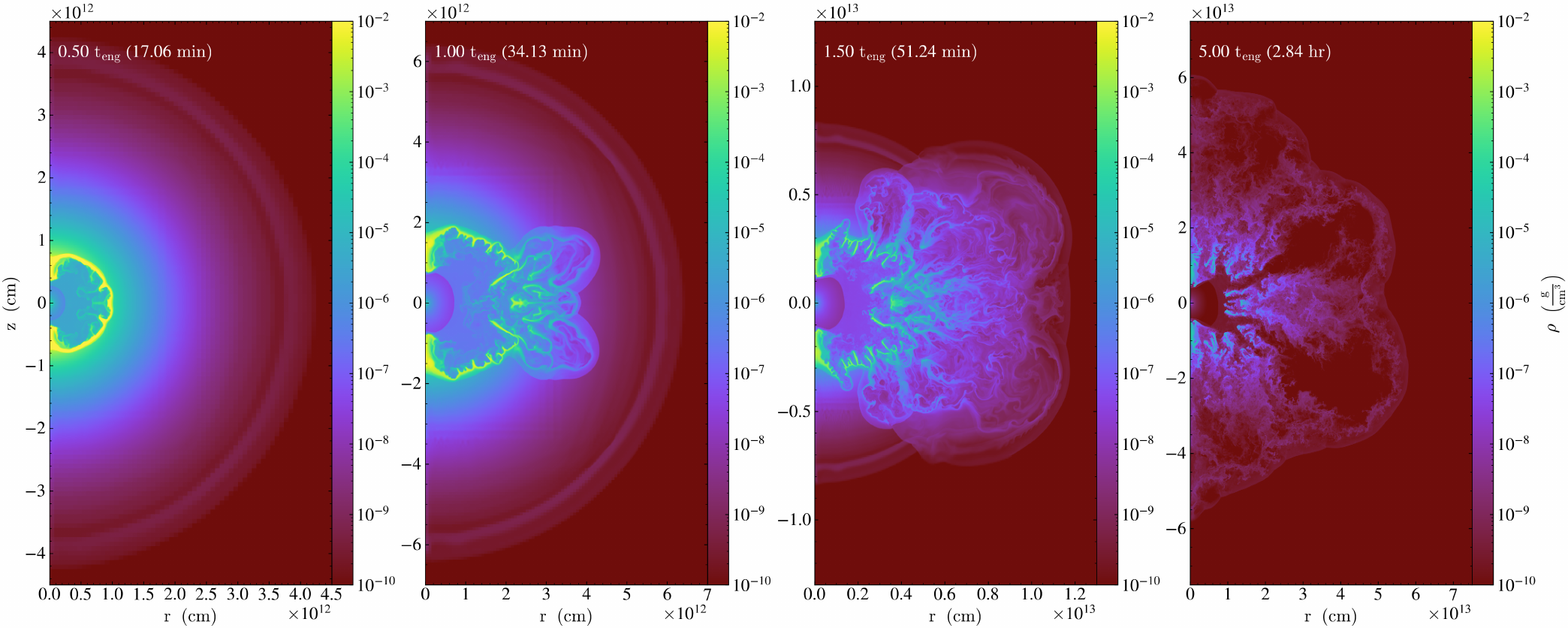}
    \caption{Density maps showing the time-evolution of the \textsf{1ms\_sin} anisotropic 
        deposition model.}
    \label{fig:sin_dyn}
\end{figure*}

\begin{figure*}
    \centering
    \plotone{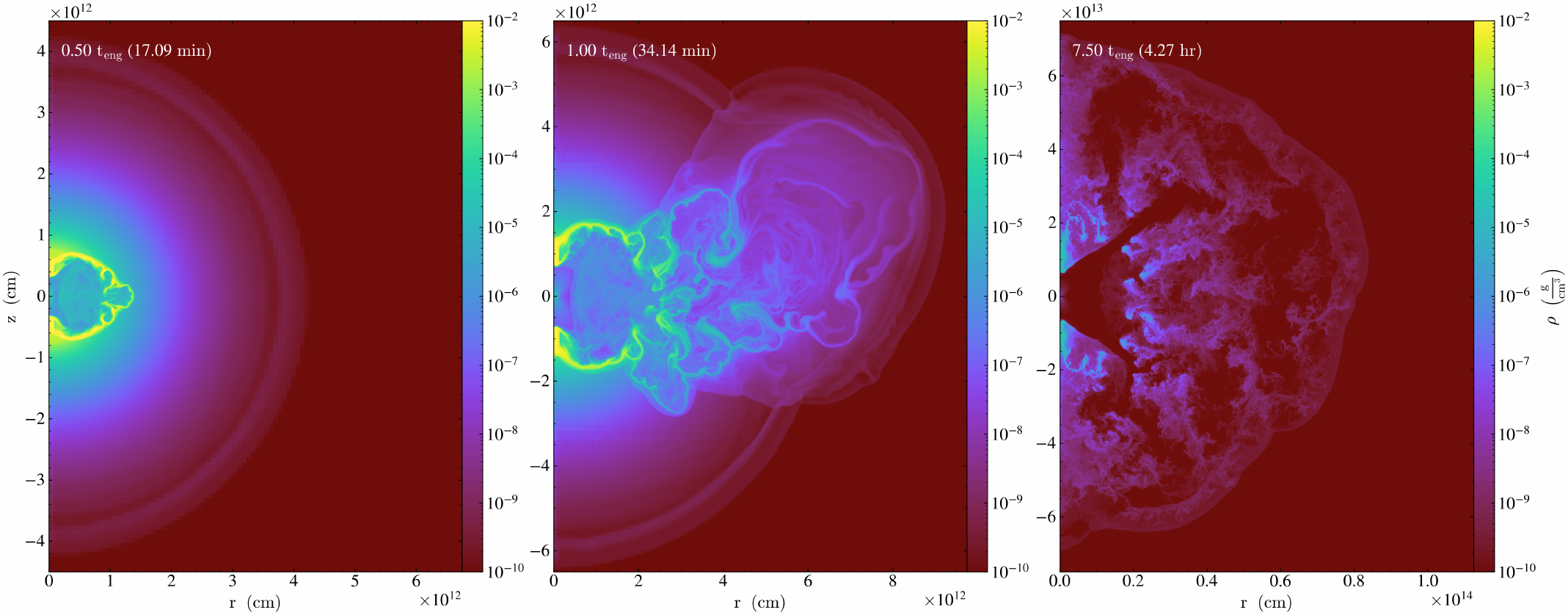}
    \caption{Density maps showing the time-evolution of the \textsf{1ms\_eq} anisotropic 
        deposition model.}
    \label{fig:eq_dyn}
\end{figure*}

\begin{figure*}
    \centering
    \plotone{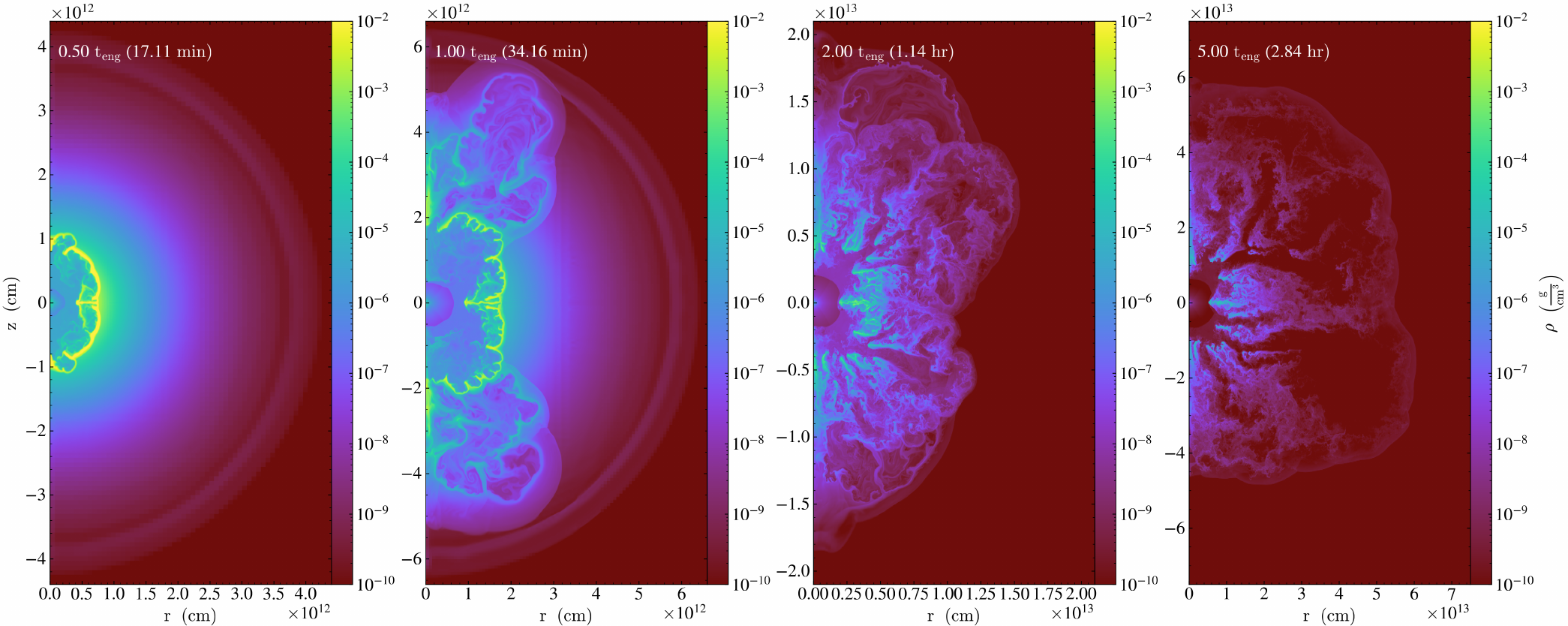}
    \caption{Density maps showing the time-evolution of the \textsf{polar\_5/3} anisotropic 
        deposition model.}
    \label{fig:bh_dyn}
\end{figure*}

Time sequences of our three simulation runs with anisotropic energy deposition
are provided in Figures \ref{fig:sin_dyn}, \ref{fig:eq_dyn} and \ref{fig:bh_dyn}.
The deposition schemes were discussed in Section \ref{sec:central_src} and
are also summarized in Table \ref{tab:sim_par}.

For our \textsf{1ms\_sin} run (Figure \ref{fig:sin_dyn}), energy is preferentially
deposited along the equatorial direction (deposition rate proportional to $\sin^2\theta$),
causing the shell to elongate. Although the
gas eventually breaks through the bubble at a similar range of angles to the
\textsf{1ms} run, it first emerges near the equatorial plane rather
than along the vertical axis. The initial breakout occurs at $t =
\numrange{0.75}{0.90}~\teng$; this is substantially later than the first breakout
in the isotropic case. This may be due to an increase in vortical (non-radial) fluid
motions in the cavity, reducing the ram pressure. The bubble ruptures at additional
locations at times in the range $\numrange{1}{2}~\teng$. Venting of
gas at later times near the sites of the original equatorial breakout
leaves behind a pair of wide, low-density channels in the remnants.

In the \textsf{1ms\_eq} run (Figure \ref{fig:eq_dyn}), where the energy is deposited
within an angle of $\pm 5\degree$ from the equatorial plane, the shell again elongates
along the $r$-axis and ruptures near the equator. The initial breakout occurs at a
similar time to the breakout in the \textsf{1ms} run ($t =
\numrange{0.45}{0.60}~\teng$), and consists of
one central outflow in the radial direction and two more that are nearly parallel
to the vertical axis. The structure of the disrupted ejecta is more sensitive
to numerical error here, with the central outflow clearly breaking from bilateral
symmetry. As in the \textsf{1ms\_sin} run, additional outflows at later times
produce more prominent channels in the remnants than were observed in the isotropic
case.

Due to preferential energy deposition in the polar direction, with the deposition rate
proportional to $\cos^2 \theta$, the \textsf{polar\_5/3} run (Figure \ref{fig:bh_dyn})
exhibits an initial breakout along the vertical axis at a similar time to the initial
breakout in the \textsf{1ms} run $t = \numrange{0.45}{0.60}~\teng$. Subsequent breakouts
occur closer to the equatorial plane. A region of denser material consisting of particles
from the engine wind and partially-disrupted
shell fragments survives in a torus surrounding the central cavity. The outflows that emerge above
and below the torus expand towards the equatorial plane and meet behind it, producing some acceleration
in the radial direction and again leaving behind wide channels in the ejecta. Despite the luminosity
decay index having a smaller magnitude than in our other simulations, the time of the initial breakout
from the cavity is comparable to that in the \textsf{1ms} simulation, and secondary outflows emerge at
similar times to those in the \textsf{1ms\_sin} and \textsf{1ms\_eq} runs.

As we have considered anisotropic engines with relatively broad effective opening angles, the morphology of the final 
remnant is not dramatically different than that of the isotropic case. One key difference in the remnant structure may 
be the presence of slower, denser ejecta at angles where the energy deposition was lower. For more collimated 
deposition, such as a jet \markchange{(see, e.g., \citealt{Chen2017})} or the equatorial outflows of \cite{Dupont2023},
the morphological differences will be more dramatic.

\section{Observable Properties}\label{sec:observables}
    
To generate synthetic observables of the models, we post-process the hydrodynamical models using the time-dependent \sedona\ Monte Carlo radiation transport code. By the last hydrodynamical time step at time $t_{\rm max}$, the ejecta velocity structure of our models is close to homology, and we use the homologous scaling laws ($r \propto t$, $\rho \propto t^{-3}$, $T \propto t^{-1}$) to remap the system to a start time of $t_0 = 1\unit{d}$ for beginning the transport calculation. This remapping assumes that pressure forces and radiation diffusion are unimportant between $t_{\rm max}$ and $t_0$, which only holds if the  engine time scale $t_{\rm eng} \ll t_0$ and effective diffusion time $t_{d} \gg t_0$. However, even when these conditions do not hold, the ejecta structure may qualitatively resemble that of our pure hydrodynamical calculation and so be sufficient to explore how the geometrical effects of engine driven dynamics influence the light curves and spectra.

When mapping the composition from our tracer elements to nuclides for post-processing, we assume that the 
progenitor was a stripped star, with an outer envelope composed of $\helium$ contaminated with a solar metallicity. \markchange{The composition of the wind (mass fraction $X_{\rm wind}$) is not known and depends on the nature of the engine, and we assume it is composed of the lightest element in our radiative transfer models ($\helium$). We discuss some consequences of this assumption in Section \ref{sec:narrow_lines}}. We convert the tracer element comprising the original intermediate shell (mass fraction $X_{\rm ims}$) to a blend of $\oxygen$ and $^{24}\mathrm{Mg}$, with a $90$ per cent to $10$ per cent split in favour of $\oxygen$, and we convert the element originally corresponding to the inner ejecta ($X_{\rm core}$) to a blend of $\nickel$ ($20$ per cent), $^{28}\mathrm{Si}$ ($79.2$ per cent), and $^{40}\mathrm{Ca}$ ($0.8$ per cent). The resultant $\nickel$ mass is $0.1\unit{M_\odot}$.

To simulate the continued energy input from the central engine, we emit radiation from a central source according to the luminosity of Equation \ref{eq:lum_eng}. In the case of an anisotropic engine, we modify the angular dependence of source emission accordingly. The exact spectrum of radiation emerging from the central engine and its surrounding nebula is uncertain, but likely non-thermal.

As a rough approximation, for the light curves and some spectra, we emit a spectrum $L_\nu \propto \nu^{-1}$ over the energy range of $\numrange{0.1\unit{keV}}{10\unit{MeV}}$. Such an approach allows us to explore the degree to which source photons are absorbed and reprocessed into optical photons in the ejecta, at least under the assumption of LTE ionization. For the remaining spectra, we use a $10,000\unit{K}$ blackbody as the central source, which should give us accurate spectral features when the optical depth is high and high-energy photons are readily reprocessed into optical. We compare the results of the two approaches in Section \ref{sec:spectra}. We also include radioactive emission from the $\nickel-^{56}\mathrm{Co}-^{56}\mathrm{Fe}$ decay chain.

The radiative transfer calculations include opacities due to Compton scattering, free-free, bound-free and lines, with the last of these treated in the expansion opacity formalism \citep{Karp1977,Eastman1993}. The gas 
ionization/excitation state are calculated assuming local thermodynamic equilibrium (LTE), with the gas temperature calculated at each timestep by balancing radiative heating and cooling. At early times, when the source photons are fully absorbed and thermalized in the ejecta, the radiation field will approach a blackbody and the LTE assumption is reasonable.

\subsection{Diffusion of Photon Packets Through the Ejecta}\label{sec:diffusion_thru_channels}

\begin{figure*}
    \centering
    \plotone{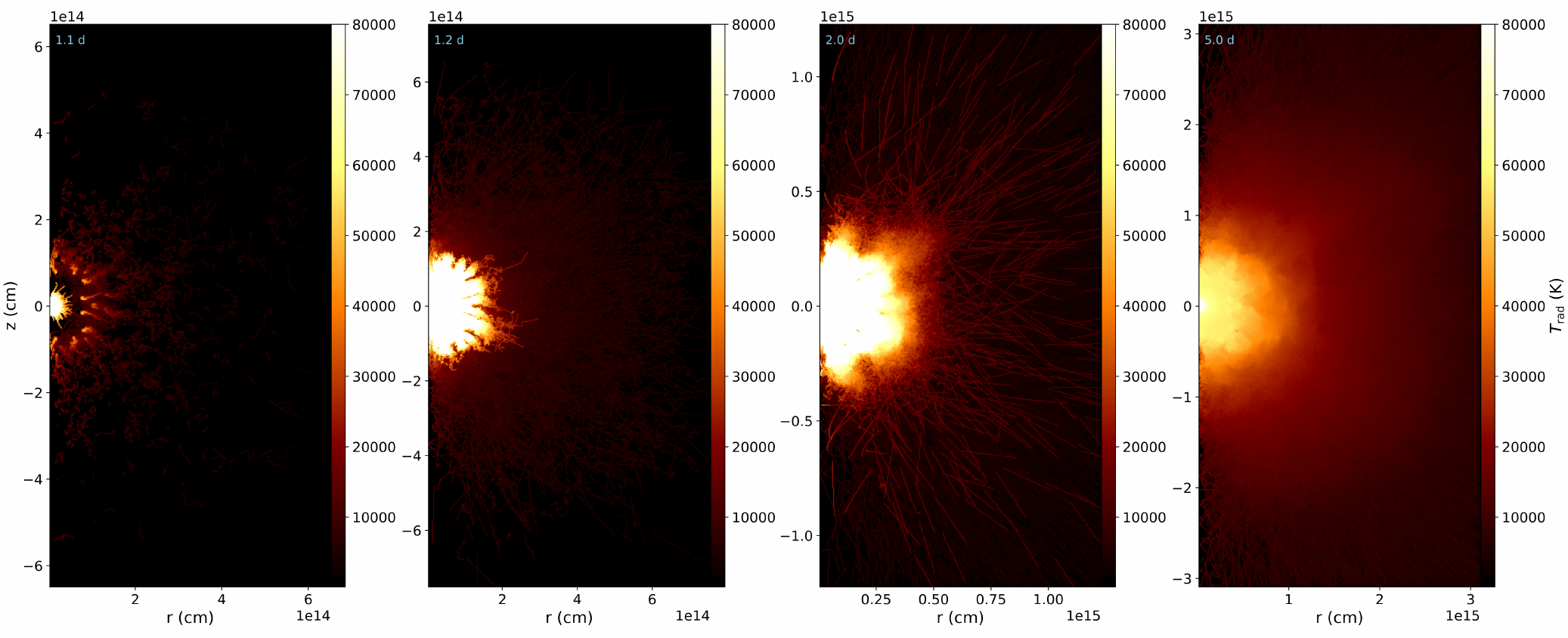}
    \caption{Radiation temperature at 4 different time points during the \sedona\ post-processing step of our 
        \textsf{1ms}. Note that the \sedona\ simulation is initialized at $t_0 = 1\unit{d}$ (just before our first 2 panels) and we are not including thermal emission when initializing the radiation field. These plots show how the concentration of Monte Carlo photon packets emitted by the central source and from $\nickel$ decay evolves with time.}
    \label{fig:sedona_t_rad}
\end{figure*}

Before discussing the observables, we examine the evolution of the radiation field in our \sedona\ post-processing step. In Figure \ref{fig:sedona_t_rad}, we plot the radiation temperature in the domain for the \textsf{1ms} run at 4 different time points, ending with $t = 5\unit{d}$ (roughly the time of peak light). We can see from the first panel that at $t = 1.1\unit{d}$, shortly after the start of the \sedona\ simulation, photon packets are streaming from the central source and are mid-way across the central cavity. There is also some radioactive emission due to $\nickel$ decay concentrated in the shell fragments bordering the low-density channels created by the wind breakout. By $t = 1.2\unit{d}$, the packets from the central source have reached the edge of the cavity and are progressing preferentially along the low-density channels. This supports the idea that these channels help facilitate the escape of radiation from the ejecta. As the denser regions of the ejecta thin, photons begin to leak out of the channels, the radiation field approaches a more isotropic configuration.

\subsection{Light Curves}

\begin{figure}
    \centering
    \plotone{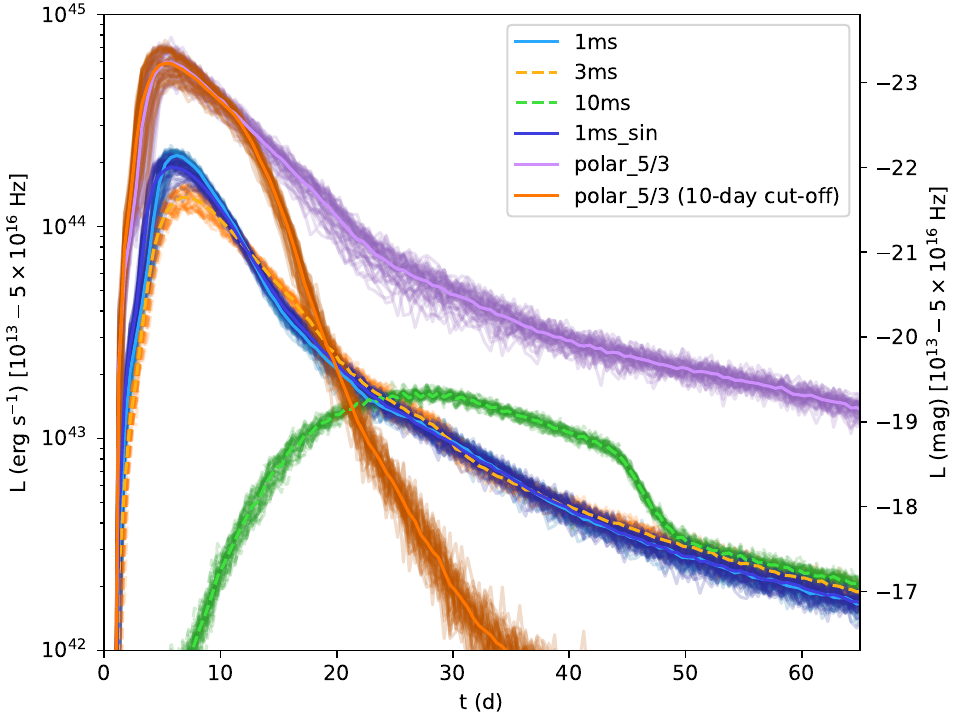}
    \caption{Synthetic light curves for several different simulation runs, where we vary the engine properties and 
    injection scheme across the runs. A summary of simulation parameters can be found in Table \ref{tab:sim_par};
    for the curve labelled ``\textsf{polar\_5/3} (10-day cut-off)" we abruptly terminate emission from the engine at
    $t = 10\unit{d}$, but it is otherwise the same as the standard \textsf{polar\_5/3} model. 
    The lighter-coloured lines in the foreground show the angle-averaged light curve, while the translucent lines in 
    the background show the angular spread. A dashed line indicates that the end time of the hydrodynamics 
    simulation exceeds the start time of the radiation transport calculation.}
    \label{fig:light_curves}
\end{figure}

We plot the light curves for 5 different simulation runs in Figure \ref{fig:light_curves}, omitting gamma rays from the luminosity. The simulations have different engine energies ($\Eeng$) and injection time-scales ($\teng$) corresponding to differing magnetar initial spin periods. Two of them (\textsf{1ms\_sin} and \textsf{polar\_5/3}) have anisotropic energy deposition schemes, and one simulation (\textsf{polar\_5/3}) uses a $k = 5/3$ engine luminosity decay instead of $k = 2$. In the case of the fast-evolving engines, we may underestimate the pre-peak luminosities, as we do not account for the radiation energy emitted prior to the start time of the transport simulations.

For runs with energetic ($\Eeng > \Ekej$), short time-scale ($\teng \ll t_d$) engines, we observe rise times of 
$\numrange{5}{10}$ days. These are consistent with the rise times of FBOTs but shorter than \markchange{those} of most SLSNe\markchange{ aside from some rapidly-evolving events such as iPTF 16asu \citep{Whitesides2017}, SN 2018gep \citep{Ho2019}, and SN 2021lwz \citep{Poidevin2025}.} Peak luminosities for these runs are $> 10^{44}\unit{erg~s^{-1}}$, consistent with both luminous FBOTs and SLSNe. The fast rise times are due to (a) the engine accelerating the ejecta and reducing the effective diffusion time and (b) the tearing of the ejecta and escape of photons through low-density channels discussed in Section \ref{sec:diffusion_thru_channels}. 1D models that do not account for these multidimensional dynamical effects may predict significantly different light curves. Our \textsf{10ms} run has a rise time more consistent with SLSNe ($\sim 28\unit{d}$) and a peak luminosity of $\sim 2 \times 10^{43}\unit{erg~s^{-1}}$, at the lower edge of the SLSN 
luminosity range. Due in part to the slower decay in engine luminosity, the \textsf{polar\_5/3} light curve exhibits 
the brightest peak and shallowest post-peak drop-off. We did not model any engines with injection time-scales 
comparable to the effective diffusion time $t_d$, which require simultaneous treatment of the hydrodynamics and 
radiation to model accurately, but these engines could potentially produce brighter events than our \textsf{10ms} run 
while still avoiding the fast rise seen in the short time-scale engines.

While the light curves of our energetic engine models rise rapidly, they decline more slowly after peak, eventually 
following the power-law decline of the central energy deposition. The dense regions of ejecta absorb much of the 
radiation from the source and reprocess it into thermal UV/optical emission. While some source radiation can escape 
through low density channels, the covering fraction of the dense ejecta filaments is $\sim 1/2$ (see 
Figure~\ref{fig:N_vs_theta}), such that substantial reprocessing persists until late times. As the optical depth drops, 
gamma-rays and hard X-rays from the source increasingly escape, but soft X-ray / far UV photons continue to be absorbed 
due to high photoionization opacities, keeping the thermalization efficiency relatively constant out to late times. 
Inclusion of NLTE photoionization may eventually enhance soft X-ray escape if the source is strong enough to completely ionize the ejecta.

In general, we expect engines with longer time-scales ($\teng \sim t_d$) to produce light curve rise times more 
consistent with SLSN (see also \citealt{Suzuki2021b}), while engines with large energy reservoirs ($\Eeng > \Ekej$) and 
comparatively short time-scales will produce rise times more characteristic of FBOTs. However, for most models, the 
post peak decline rate is much slower than that observed in luminous FBOT light curves. The exception is the polar 
model for which we shut off the source after $10$ days, resulting in a rapid drop in the bolometric luminosity similar 
to that seen in luminous FBOTs. The light curve after the engine shutoff represents the continued diffusion of the 
remaining radiation trapped in the optically thick ejecta. Such a scenario might represent a black hole central engine 
in which engine activity unbinds the infalling material and eventually cuts off accretion. Additionally, the rate of 
drop off in the luminosity depends on the thermalization efficiency of high-energy source photons in the ejecta. 
Reducing the ejecta mass or including NLTE effects could also facilitate the escape of source photons at later times 
and hasten the decline of the light curve.

Figure \ref{fig:light_curves} also shows the angular variation of the light curve for each simulation. We find only minor variation in brightness with viewing angle (about $\numrange{10}{20}$ per cent) for our runs with isotropic energy injection, and there is no qualitative difference in the light curve shape. This is because the observed signal is a sum of the emission over the projected area of the system, such that the smaller scale asymmetries caused Rayleigh Taylor instabilities are washed out. Even in calculations with an anisotropic source (e.g. \textsf{polar\_5/3}), the brightness varies with angle at most a factor $\numrange{3}{5}$ at any given time. This is primarily due to the assumed angular dependence of the source, and not the asymmetry of the ejecta itself.

\begin{figure}
    \centering
    \plotone{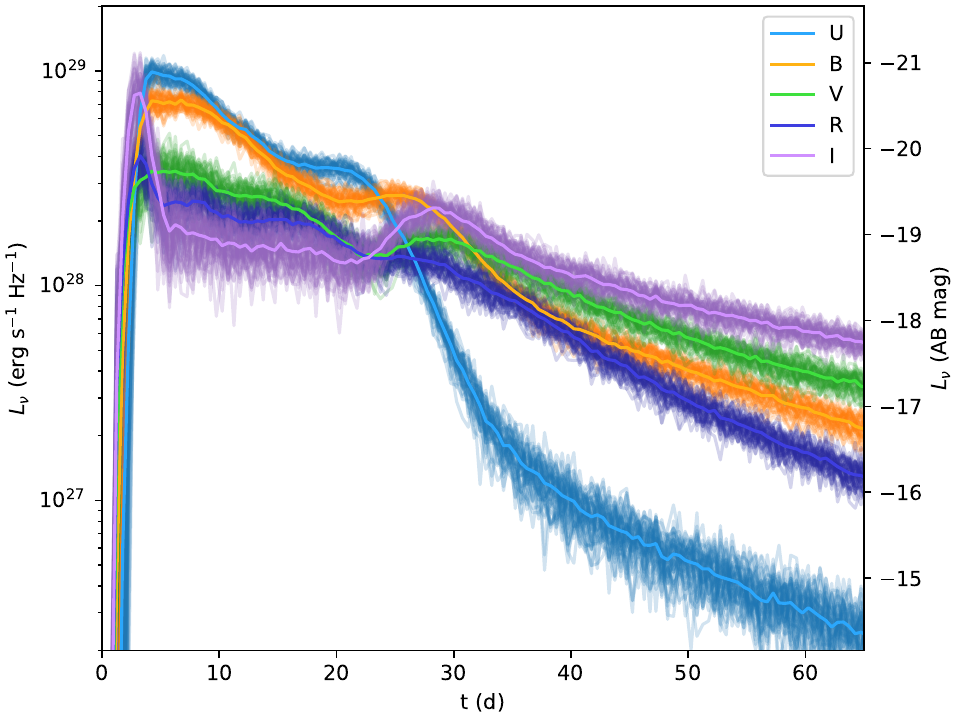}
    \caption{Synthetic light curves for our \textsf{1ms} simulation run in five different photometric bands (Johnson U, B, and V, Cousins R and I). The y-axes are bandpass-averaged luminosity (left) and AB magnitude at $10\unit{pc}$ (right). As in Figure \ref{fig:light_curves}, the lighter-coloured lines in the foreground show the angle-averaged light curve, while the translucent lines in the background show the angular spread.}
    \label{fig:light_curves_band_comp}
\end{figure}

\begin{figure}
    \centering
    \plotone{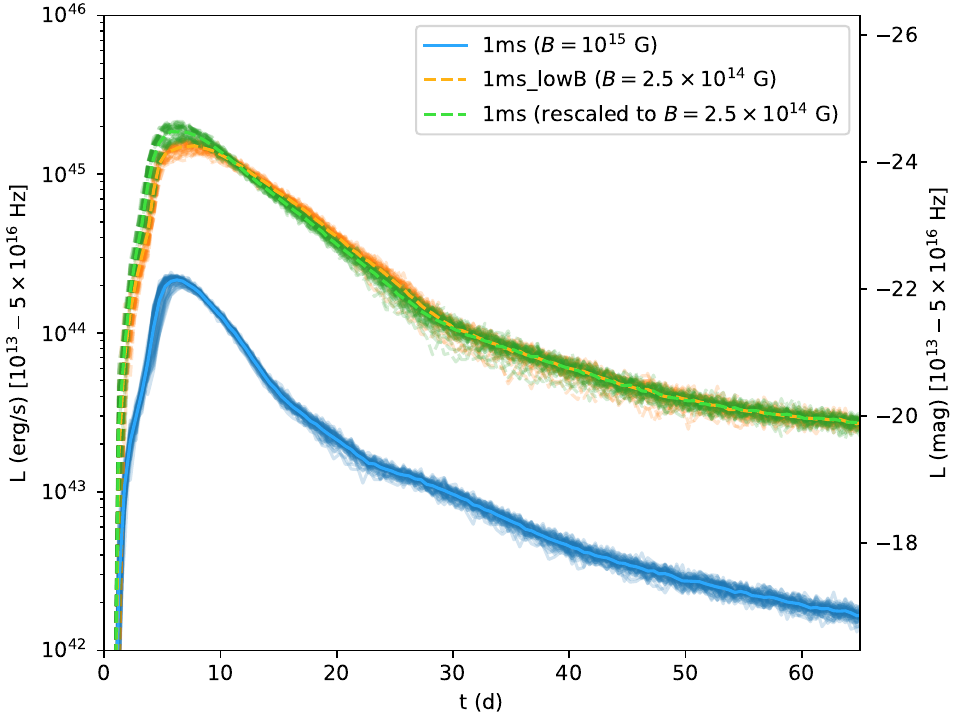}
    \caption{Synthetic bolometric light curves for our simulations where we vary only the engine energy injection 
    time-scale (magnetic field for a magnetar engine) across the runs. Two of the models (\textsf{1ms}, 
    \textsf{1ms\_lowB}) are listed in Table \ref{tab:sim_par}, and correspond to magnetar engines with $B = 
    10^{15}\unit{G}$ and $B = 2.5\e{14}\unit{G}$ respectively. The other light curve (also corresponding to a $B = 
    2.5\e{14}\unit{G}$ magnetar) was obtained by rescaling the output of our \textsf{1ms} calculation using the 
    transformation to dimensionless space discussed in Section \ref{sec:dim_par}, and using that to initialize the 
    \sedona\ calculation. The lighter-coloured lines in the foreground show the angle-averaged light curve, while 
    the translucent lines in the background show the angular spread. A dashed or dash-dotted line indicates that 
    the end time of the hydrodynamics simulation exceeds the start time of the radiation transport calculation.}
    \label{fig:light_curves_timescale_comp}
\end{figure}

Figure \ref{fig:light_curves_band_comp} shows broadband light curves of the \textsf{1ms} simulation. In this model, the emission around peak is primarily in U and B-bands, but evolves toward the red over time. The V and R-band light curves show a more gradual decline as the redward shifting of the spectral energy distribution compensates for the declining bolometric luminosity. The broadband light curves show secondary peaks, which arise from changes in ionization state and the onset of line blanketing as the ejecta cool, similar to what is seen in Type Ia supernovae due to changes in ionization state of the iron group elements \citep{Kasen2006a}. Such radiative transfer effects could contribute to the bumpiness observed in some SLSNe, though the effect will be model dependent. In the \textsf{1ms} model of Figure \ref{fig:light_curves_band_comp}, the engine mixes iron group elements into the outer layers of ejecta, which enhances their effect on the spectral energy distribution. For weaker engines with less mixing, or models that produce little $^{56}$Ni, secondary light curve bumps would be less pronounced or absent.  

We tested the temporal rescaling described in Section \ref{sec:dim_par} by running and post-processing the 
\textsf{1ms\_lowB} simulation and comparing the resultant light curve to one obtained by rescaling our \textsf{1ms} 
run. Our \textsf{1ms\_lowB} simulation uses a magnetar engine model with an effective B-field of $2.5 \times 
10^{14}\unit{G}$, which results in an engine time-scale $16$x longer than that of the \textsf{1ms} simulation. We 
should be able to simulate the effect of the extended engine time-scale by transforming to the dimensionless space 
defined by the unit system of Section \ref{sec:dim_par}, and transforming back into the dimensionful space assuming 
a different engine time-scale. Using this technique, we obtained a rescaled initial model for the \sedona\ 
post-processing step (using the longer time-scale for the \sedona\ radiation source as well), and compare the 
results in Figure \ref{fig:light_curves_timescale_comp}. The \textsf{1ms\_lowB} and rescaled models show good 
agreement, with some minor divergence caused by a small amount of structural variation.

\subsection{Spectra}\label{sec:spectra}

\begin{figure}
    \centering
    \plotone{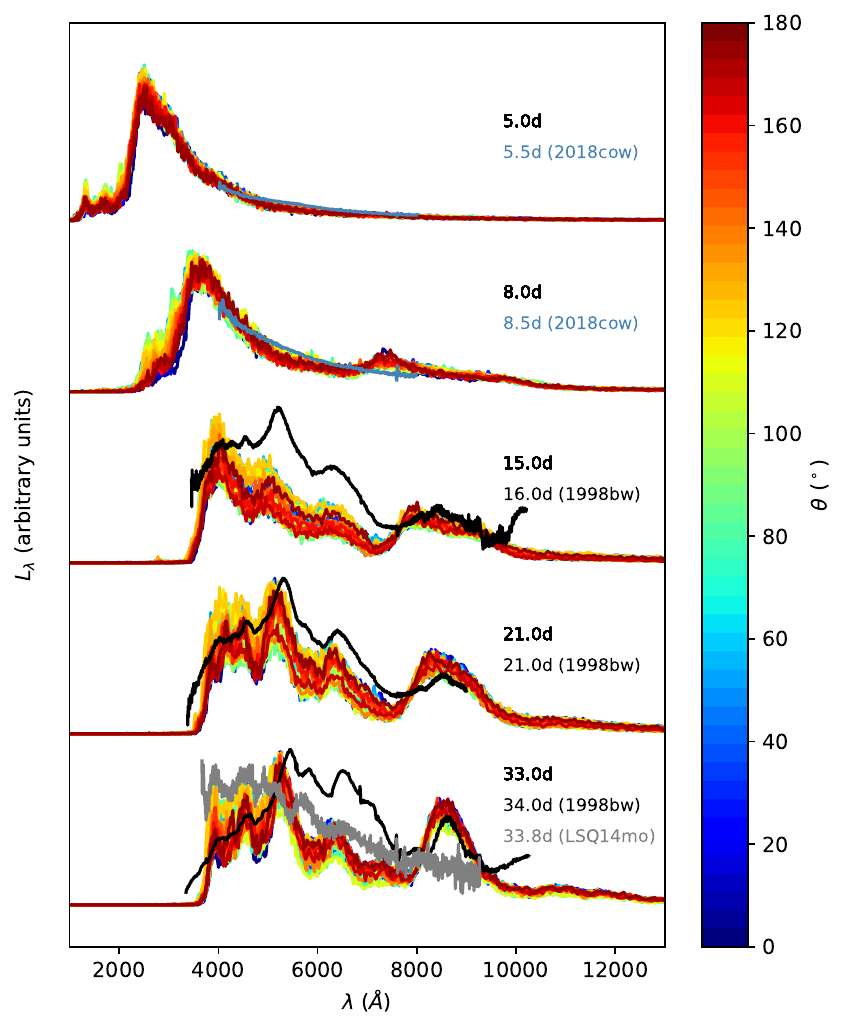}
    \caption{Synthetic spectra for \textsf{1ms} simulation run at a sequence of time points (given by the
    topmost label in each stack of labels, measured from the estimated time of explosion). For these spectra we
    used a $10,000\unit{K}$ blackbody as the central source. The spectra are plotted at a range of viewing angles, with the polar angle given by the line colour. Spectra at the closest time point for SN 1998bw (black; data from \citealt{Patat2001}), SN 2018cow (blue; data from \citealt{Prentice2018}) and SN LSQ14mo (dark gray; data from \citealt{Chen2017_LSQ14mo}) are plotted over some of the simulated spectra for comparison.}
    \label{fig:spectra}
\end{figure}

We generate higher signal-to-noise synthetic spectra from our models by performing iterative ``steady-state" calculations at particular time points. We first examine the models with a blackbody central source. A spectral time series for our \textsf{1ms} model with a blackbody source is given in Figure \ref{fig:spectra}, where we also overplot the spectra of an SN Ic-BL (SN 1998bw), an FBOT (SN 2018cow), and a Type I SLSN (SN LSQ14mo) for comparison. We emphasize that the models are not tuned to produce spectra similar to the comparison events, and we include the comparison spectra mainly to highlight the broad similarity in features at various epochs. 

At early times ($t \lesssim 10\unit{d}$ since time of explosion) the spectra are hot and approximately blackbody, \markchange{and lack the spectral features characteristic of SLSNe before or around peak light. These include the archetypal W-shaped \ion{O}{II} absorption feature from approximately $4000$ to $4500$ \AA\ present in the pre-peak spectra of many SLSNe \citep{Quimby2011,Konyves-Toth2021,Konyves-Toth2022}, and frequently seen C, Si, and Fe lines. The spectra at these times instead} resemble the featureless spectra of FBOTs, or \markchange{the similarly featureless near-peak spectra of the rapidly-evolving SLSN Ic-BL iPTF 16asu \citep{Whitesides2017}. The featurelessness in our synthetic spectra is likely a result of the high source luminosity, which ionizes the surrounding ejecta and inhibits the development of line features. Doppler broadening due to fast expansion velocities may also help lines blend into the continuum \citep{Aspegren2026}.}

By $t \sim \numrange{10}{20}\unit{d}$ since explosion, the ejecta have cooled and spectral features have begun to 
emerge. The line profiles are broad, reflecting the high-velocities caused by the acceleration of the ejecta by the 
engine. \markchange{This is in tension with luminous FBOTs, whose spectra often remain largely featureless for many 
weeks after detection \citep{Margutti2019,LeBaron2026}.} The spectral region $\numrange{5000}{9000}$ \AA\ is shaped by 
line features of Ca \markchange{(the hump at $\numrange{8000}{9000}$ \AA)}, O \markchange{(at $\sim 7800$ \AA)}, and Si 
\markchange{(at $\sim 6000$ \AA)}, which qualitatively resemble those of an SN Ic-BL such as SN 1998bw. \markchange{A 
spectral transition from a featureless continuum to a spectrum more resembling an SN Ic-BL is also seen in the event 
iPTF 16asu \citep{Whitesides2017}.}

Due to the higher luminosity, the model is bluer than SN 1998bw, and the model colours at $\sim 33\unit{d}$ are closer 
to that of the brighter SLSN event SN LSQ14mo. \markchange{The model achieves best agreement with the comparison event 
SN LSQ14mo in the $\sim\numrange{4000}{5000}$ \AA\ range, where post-peak SLSN spectra tend to be dominated by Ca, Mg, 
and Fe features \citep{Gal-Yam2012,Chen2017_LSQ14mo,Aamer2025}. These include possible contributions from \ion{Ca}{II} 
at $\sim 4000$ \AA, \ion{Mg}{I} at $\sim 4600$ \AA, and \ion{Fe}{III} at $\sim 5000$ \AA. While not visible in the 
spectra of LSQ14mo, the peak at $\sim 5300$ \AA\ (attributed to Fe) is also commonly seen in SLSNe by $20$ to $40$ days 
post-peak \citep{Aamer2025}. There is a sharp decrease in flux blueward of $\sim 4000$ \AA, likely due to line 
blanketing from fast-moving Fe-group elements that were mixed outward by the breakout of the wind from the cavity (see 
also the discussion of the spectra from the \textsf{10ms} run below). This is not typical of SLSN. However, significant 
absorption in the near-UV is also seen in near-peak and post-peak spectra of some SNe Ic-BL 
\citep{Iwamoto2000,Nomoto2001,Foley2003}, and in SN 2017dwh \citep{Blanchard2019}, which is more similar to an SLSN 
than an SN Ic-BL near peak but its spectra evolve post-peak to more closely resemble the latter.} By tuning the model 
parameters, it is likely we could achieve better agreement with observed \markchange{SLSN or SN Ic-BL} spectra.

With an isotropic source, there is little dependence on viewing angle in the spectra, although a relatively small angular spread emerges by $\sim 10\unit{d}$ post-peak, e.g., around $6000$ \AA. Whatever angular variation there is arises from the formation of low-density channels and the anisotropy in chemical composition caused by the wind breakout.

\begin{figure}
    \centering
    \plotone{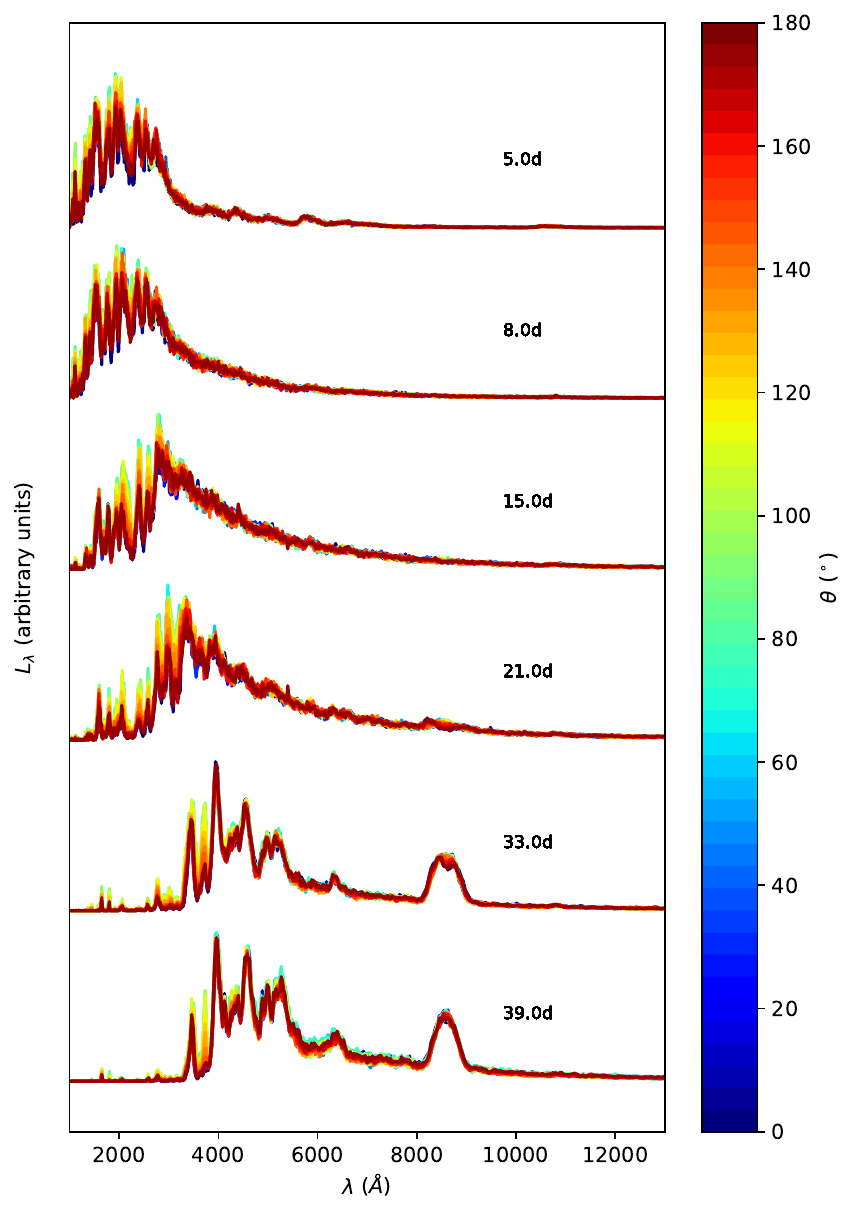}
    \caption{Synthetic spectra for \textsf{10ms} simulation run at a sequence of time points (given by the
    topmost label in each stack of labels, measured from the estimated time of explosion). For these spectra we
    used a $10,000\unit{K}$ blackbody as the central source. The spectra are plotted at a range of viewing angles, 
    with the polar angle given by the line colour.}
    \label{fig:spectra_10ms}
\end{figure}

This evolution from an FBOT-like spectrum to one more consistent with broad-lined SLSNe may be characteristic of 
moderate-mass ejecta harboring engines with large energy reservoirs ($\Eeng \gtrsim \Ekej$) and short injection 
time-scales ($\teng \ll t_d$. \markchange{The estimated ejecta velocities in SLSNe tend to occupy the lower end of the 
SNe Ic-BL velocity distribution \citep{Aamer2025}, and they typically exhibit less extreme Doppler broadening than seen 
in Figure \ref{fig:spectra}. Extreme broadening can be avoided with a weaker engine that does not substantially 
accelerate the ejecta (see also the discussion of Figure \ref{fig:spectra_10ms} in the next paragraph)}, or perhaps by 
increasing the injection time-scale of a stronger engine to the point that radiative losses sap enough energy to 
forestall the wind breakout. \markchange{A more FBOT-like long-term spectral evolution with longer time-scales for 
spectral features to emerge can potentially be achieved} by varying the ejecta mass, ejecta density structure, and 
engine luminosity curve.

Figure \ref{fig:spectra_10ms} gives a spectral time series for our \textsf{10ms} simulation, which features a weaker 
engine that is unable to rupture the central bubble after its formation. Due to the lack of a breakout, there is no 
substantial acceleration of the outer ejecta and there is no large-scale restructuring to enhance anisotropy in column 
depth and composition, \markchange{and there is significantly less outward mixing of heavy elements}. The resulting 
spectra show narrower lines, less variation with viewing angle, \markchange{and less line blanketing in the near-UV 
(caused by broad Fe-group absorption features) than those of the \textsf{1ms} model. Like in the \textsf{1ms} case, the 
W-shaped feature present in many SLSNe prior to peak is absent here, which may be due to this model having too low an 
engine luminosity and photospheric temperature, or the lack of an NLTE treatment in the spectral calculations (see 
\citealt{Konyves-Toth2022,Saito2024}).} The reduced velocity compared to the \textsf{1ms} also results in slower 
spectral evolution, with features just starting to appear in the optical $\numrange{15}{20}$ days post-explosion or 
$\numrange{5}{10}$ days before peak light. The spectra at this phase resemble those of normal Type Ic supernovae.

\subsubsection{Possible Narrow Lines from a Central Engine}\label{sec:narrow_lines}

\begin{figure}
    \centering
    \plotone{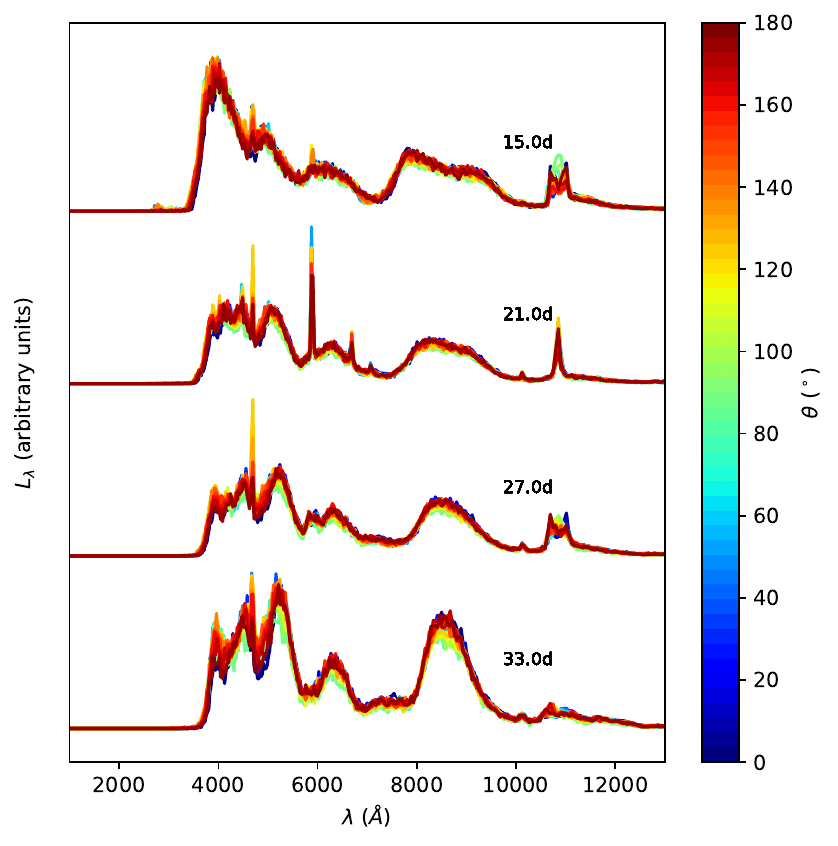}
    \caption{Synthetic spectra for \textsf{1ms} simulation run at a sequence of time points (given by the
    topmost label in each stack of labels, measured from the estimated time of explosion). For these calculations we had the central source injecting high-energy ($> 0.1\unit{keV}$) photons with a power-law input spectrum. The spectra are plotted at a range of viewing angles, with the polar angle given by the line colour.}
    \label{fig:spectra_1ms_nt}
\end{figure}

While we assumed in our first set of spectral calculations that the central engine radiates as a blackbody, the true emission spectrum may be non-thermal. For example, dissipation of energy at a wind termination shock or due to magnetic reconnection in the bulk can accelerate non-thermal electrons which will radiate through synchrotron and free-free processes, likely producing a power-law photon spectrum \citep{Arons2003,Vurm2021}.

As a rough approximation of a non-thermal source, we run additional spectral calculation of the 1ms model assuming the central source emits a power-law spectrum  $L_\nu \propto \nu^{-1}$ over the energy range of $\numrange{0.1\unit{keV}}{10\unit{MeV}}$. While the ionization/excitation state of gas directly irradiated by such a source is likely to depart significantly from LTE, we none the less retain the LTE assumption and explore the qualitative behaviour. LTE is likely to remain a reasonable approximation in optically thick regions where the source spectrum is efficiently absorbed and reprocessed into thermal photons.  

Figure~\ref{fig:spectra_1ms_nt} shows the spectral evolution of a model with a non-thermal source. X-rays are largely absorbed through photoionization and reprocessed in optical thick regions into thermal optical photons. Overall, the optical light curves and spectra  closely resemble the model using a blackbody source. A key difference is the appearance of narrow He emission lines, which arise from the reprocessing of high energy photons by material in the central cavity. These lines are strongest around 20 days, when the ejecta optical depth has become low enough that the line photons can escape from the central region. They fade as the source decreases in luminosity and the density of the inner region drops.

\markchange{The appearance of these narrow lines is only suggestive. As our calculations assume LTE}, we do not capture the realistic photoionized conditions in the cavity, which may diverge significantly from LTE. Our homologous models may misrepresent the interior velocity structure, as they enforce low velocities at small radii. \markchange{The lines also require helium to be present in the cavity; in this case the helium comes from the assumption that the engine wind is composed of helium. The approximation of a helium-rich wind is not completely unreasonable, as helium can potentially persist in fast outflows from neutron stars due to alpha-rich freeze-out. However, a wind with a different composition and mass loading than used here may not introduce enough helium to the cavity to produce the He emission lines, in which case helium would need to already exist in the vicinity of the engine for those lines to appear. We note that a wind dominated by hydrogen could produce narrow Balmer lines, but these were not observed in test calculations with a pure hydrogen wind for the luminosities and wind mass loading studied here. Metal-rich winds might also result in visible metal lines, but further study is needed to determine which metal lines may appear, if any.}

\markchange{Working under the assumption of a pure helium wind composition,} we can analytically estimate the physical conditions that could lead to narrow line emission. For a fully ionized medium, the emissivity (units ${\rm erg~s^{-1}~cm^{-3}}$) of the $4686$ \AA\ \ion{He}{II} line is 
\begin{equation}
    j_{4686} = n_e ~n_{\rm HeII}~ \alpha_{\rm B, HeII}~ f_{4686} E_{4686} 
\end{equation}
where $n_e$ is the free electron density, $\alpha_{\rm B, HeII} \approx 1.5 \times 10^{-12}~{\rm cm^3~s^{-1}}$ is the recombination coefficient for recombination to excited states of \ion{He}{II}, $f_{4686} \approx 0.12$ is the fraction of recombination cascades that produce a $4686$ \AA\ photon, and $E_{4686}$ is the energy of a $4686$ \AA\ photon. This expression assumes that the source is not so strong that the excited states of \ion{He}{II} are photoionized before they can cascade down and produce line photons. The implied total luminosity of the $4686$ \AA\ line from a mass $M_c$ of helium moving homologously up to speed $v_c$ in a spherical region of radius $r_c = v_c t$ is 
\begin{align}
    L_{4686} &= \left(4 \pi r_c^3/3\right) j_{4686}\\ 
    &\approx 2 \times 10^{40}~\left(\frac{M_c}{10^{-5}M_\odot}\right)^{2} \left(\frac{v_c}{\left[5\e{-4}\right] c}\right)^{-3} \left(\frac{t}{20.0\unit{d}}\right)^{-3} \unit{erg~s^{-1}}
\end{align}
This line emission will be spread over a wavelength region of $\Delta \lambda = \lambda_0 (v_c/c)$ where $\lambda_0$ is the line rest wavelength. The continuum luminosity, $L_{\rm bol}$, is roughly distributed over an optical wavelength band of width $\sim \lambda_0$, so the line emission will be conspicuous  when $L_{\lambda,4686} / L_{\rm bol}  \gtrsim v/c$. This suggests that a small mass ($M_c \sim 10^{-5}~M_\odot$) of material moving at low velocity ($v \approx 5\e{-4}c$) inside the cavity could produce noticeable line emission, provided the line photons can escape without being reabsorbed and reprocessed in the overlying dense ejecta shell. This possibility should be investigated with further multi-dimensional NLTE calculations.

\section{Discussion and Conclusion}\label{sec:discussion}
    
We performed 2D hydrodynamics calculations of CCSNe powered by long-lived central engines, and used post-processing radiation transport to obtain synthetic light curves and spectra. The simulations demonstrate the importance of multidimensional effects in the ejecta dynamics and observational signatures. The results of the post-processing also suggest a connection between FBOTs, SLSNe, and SNe Ic-BL, and support the idea that central engines play a role in each of these classes of events.

\subsection{Summary of Main Results}

Energy injection from the engine inflates a high pressure bubble in the ejecta interior, producing RT instabilities
at the interface. In the limit that the energy deposition occurs early enough that the ejecta is optically thick and
radiative diffusion is negligible, the dynamical evolution is determined by the ratio of engine energy to ejecta
kinetic energy $\Eeng/\Ekej$. When this ratio $\Eeng/\Ekej \gtrsim 1$, the bubble ruptures and outflows driven
by the wind from the engine propagate through the entire ejecta. The outflows accelerate and shred the outer
ejecta and leave behind a clumpy remnant with low-density channels, where the channels have more than an order
of magnitude lower column density than the surrounding ejecta. Elements originally confined to the inner regions
of the ejecta are mixed throughout the remnant. If the engine energy reservoir is small compared to
the ejecta kinetic energy, the engine wind fails to break out of the bubble and disrupt the outer ejecta.

For regions of the parameter space where the bubble ruptures on time-scales shorter than the ejecta radiative 
diffusion time (large engine energy reservoirs, short injection time-scales), our post-processing results indicate 
that we would see a luminous, fast-evolving transient with luminosity $\gtrsim 10^{44}\unit{erg~s^{-1}}$ and a rise time on the order of days. These basic photometric properties are similar to those of luminous FBOTs, although the UVOIR light curves of our models show relatively slow decline rates compared to FBOTs unless we shut off the central engine. Spectra are initially featureless, again resembling an FBOT, but they begin to develop spectral features consistent with bluer SLSNe/SNe Ic-BL by $t \sim 20\unit{d}$, while FBOT spectra tend to remain largely featureless out to longer timescales.

Engines with longer injection time-scales can produce rise times more consistent with SLSNe 
($\gtrsim 10\unit{d}$). Magnetar engines with periods of $\sim 10\unit{ms}$ can still achieve peak luminosities of 
$\sim 10^{43}\unit{erg~s^{-1}}$, but shorter rotation periods or additional energy sources are likely required to 
produce the most luminous slow-evolving events. These results provide some evidence that central engine models can
explain at least a subset of FBOTs, SLSNe, and SNe Ic-BL, but we need to explore additional areas of the parameter space (e.g.\ longer injection time-scales, larger or smaller ejecta masses, different density structures) and include additional physics to capture the full range of possible observational signatures.

\subsection{Discussion}

The dynamical effect of a central engine has several observable implications. For strong engines ($\Eeng/\Ekej > 1$), high-pressure gas in the cavity eventually breaks out of it and vents through the rupture points. The resulting high-velocity outflows could produce prominent radio emission when they interact with the CSM. The outflows in our simulations are largely at sub-relativistic speeds ($< 0.4c$), but this value is highly uncertain, as the mass-loading of our engine wind and the density of the CSM are free parameters, the hydrodynamics is non-relativistic, and we neglect radiative losses. Nevertheless, the models suggest that the kinetic energy of these near relativistic outflows can be significant (several to ten percent of $\Eeng$) in which case they are likely to power significant radio emission. 

The absence of early-time radio detections in most SLSNe \citep{Margutti2023} may indicate that the central engine is not energetic enough for the bubble to break out ($\Eeng/\Ekej \lesssim 1$), or that it does so only after radiative losses have already drained much of the available energy. The latter would be consistent with a scenario where SLSNe are powered by long-lived engines with time-scales comparable or greater than the diffusion time. For the faster evolving, powerful engines relevant for luminous FBOTs, bright radio emission could arise through this blowout mechanism. Multiple broad outflows form from spatially separated rupture points, which could lead to distinctive features in the resulting radio light curves. 

If an engine driven shock breakout occurs diffusively, the radiated energy could produce bumps in the UV/optical light curve (e.g., \citealt{Kasen2016}), which may be relevant for explaining the early peak seen in some SLSNe light curves (e.g., \citealt{Leloudas2012,Nicholl2015}). In multiple dimensions, this breakout occurs at discrete locations and times, which could potentially lead to multiple bumps in the bolometric light curve. Full radiation hydrodynamical simulations will be needed to determine whether such a process can produce distinct, temporally resolvable features, such as those seen in many SLSNe \citep{Lunnan2018,Hosseinzadeh2022,Chen2023}. 

The restructuring of the ejecta seen in our models results in faster rising light curves compared to 1D models that neglect the multidimensional dynamics. The stochastic breakout opens up channels with column densities more than an order of magnitude below the surrounding ejecta. We find this allows optical radiation to escape more rapidly, reducing the time to peak luminosity. In addition, the engine-driven acceleration of the ejecta shortens the overall effective diffusion time, also leading to a more rapid rise. Simplified 1D treatments that neglect the multidimensional dynamics may yield biased inferences about the properties of the engine and ejecta.

The channels may also facilitate the escape of non-thermal radiation from interior regions (e.g., a magnetar 
synchrotron nebula or wind termination shock). Radio emission from the central region, which is generally obscured by 
free-free absorption in the ejecta, will be able to escape earlier through these lower-density pathways. The low column 
density will also make it easier for X-rays to fully ionize the channels and emerge \citep{Metzger2014}. An NLTE 
radiation transport treatment will be necessary to accurately capture these effects.

\markchange{The engine wind breaking out of the central cavity may play a role in explaining some events documented in the literature that are fast-rising and/or show signs of a transition from one class of event to another. Among these events are iPTF 16asu (analyzed in \citealt{Whitesides2017}), SN 2018gep \citep{Ho2019}, and SN 2021lwz \citep{Poidevin2025}. All three of these events develop spectra characteristic of SN Ic-BL after maximum light, have unusually blue spectra prior to maximum light, and have relatively short rise times compared to most SLSNe (estimated to fall in the range $\numrange{3}{8}$ days). The initial spectra of iPTF 16asu are featureless continua resembling FBOT spectra, and thus the spectral evolution is similar to that of our models where $\Eengtild > 1$ and the wind breaks out of the cavity. The early spectra of 2018gep exhibit the W-shaped \ion{O}{II} feature commonly seen in SLSNe \citep{Ho2019}. SN 2021lwz shows spectral features associated with normal Type Ic supernovae prior to peak, with lines broadening after peak.}

\markchange{A blowout caused by an engine could explain a fast rise and broad post-peak spectral features even with a relatively large ejecta mass. Energy injection from the engine would cause blue spectra, although adjustment of the engine and ejecta parameters from our models would be needed to see early-time spectra similar to those of 2018gep and 2021lwz. SN 2018gep declines very rapidly immediately after peak compared to our magnetar models; \citet{Ho2019} suggested that the early light curve in SN is well-described by shock breakout in an extended CSM, but an additional energy source is needed to explain the late-time light curve. The lack of strong radio detection in iPTF 16asu and SN 2018gep disfavors outflows with large kinetic energy continuing to interact with the environment after the initial blowout.}

\markchange{Another interesting transitional event is SN 2017dwh \citep{Blanchard2019}, a hydrogen-poor SLSN whose spectra redden with time and evolve to look like those of an SN Ic-BL. SN 2017dwh also shows evidence for Fe-group elements in its outer ejecta. As seen in our hydrodynamics models, full or partial break out of an engine wind (or possibly a jet) from the inner ejecta will disperse heavy elements throughout the outer ejecta, which may account for the presence of Fe-group elements in the outer ejecta of SN 2017dwh. This would also accelerate the ejecta, producing broad lines following the blowout. If a blowout occurs at a time comparable to or later than the peak of the light curve, it can avoid significantly shortening the supernova rise time and producing an unusually fast-evolving event. This may be necessary to reconcile the rise time of about $19$ days in SN 2017dwh with a model where the engine wind escapes the central cavity.}

Finally, the excitation of gas within the bubble cavity may also produce distinctive spectroscopic signatures. Photoionization of gas surrounding the engine can produce line photons that may escape through low-density ejecta channels with relatively little further reprocessing. If the gas in the interior is moving at relatively low velocity, this may produce narrow line emission. While these narrow lines would resemble those produced by external circumstellar interaction, they would in fact be a signature of an internal central engine.

\subsection{Future Work}

We rely on numerical error to seed instabilities during the inflation of the bubble, and even with AMR numerical effects may further alter the wind breakout pattern. Therefore, the final density structure and the variation with viewing angle in the light curves and spectra have some dependence on the numerical grid. Seeding density perturbations in the simulation setup or initializing from an existing multidimensional model would ameliorate this somewhat. Performing the simulations in 2D also alters the turbulent cascade and the development of instabilities, and 3D simulations would better capture the mixing of elements and smaller scale structures that arise from the instabilities \citep{Blondin2017,Suzuki2019,Chen2020}.

There is considerable variability and uncertainty in the engine behaviour and how the energy is deposited in the ejecta that is not captured by our simple injection model. The underlying engine behaviour is especially visible in the tails of the light curves, when the luminosity decay begins to track that of the central energy source. Using more complex models of the energy injection from the engine could enable us to reproduce photometric features that were not seen in our synthetic light curves (e.g. multiple post-peak bumps, different decay rates). An NLTE treatment of the radiation transport could also sharpen the decay in the light curves if NLTE effects allow X-rays can escape more easily.

Finally, our ability to model engines with longer characteristic time-scales is limited by the fact that we do not include any treatment of radiation in our hydrodynamics simulations. When the characteristic time-scale is comparable to the diffusion time through the ejecta, radiation may begin to leak out of the central bubble and delay or suppress the wind breakout. Including radiation should also allow us to accurately estimate the 
temperature of the ejecta and thus better model the early part of supernova light curve. Our spectral modelling out to long time-scales can also be improved by including NLTE effects in our radiation transport calculations, using a more physical radiation spectrum for our central source, and/or attempting to more accurately estimate or simulate the composition of the ejecta. We leave these improvements to future studies.

\section*{Acknowledgements}

\markchange{We thank the anonymous referee for constructive comments on the manuscript.} We thank Raffaella Margutti, 
Wenbin Lu, Peter Nugent, Ke-Jung Chen, Brian Metzger, V.\ Ashley Villar, Peter Blanchard, and Edo Berger for insightful 
discussions, Matt Nicholl for discussion and for suggesting LSQ14mo as a comparison event for our spectra, and Michael 
Zingale, Maximilian P.\ Katz, Eric T.\ Johnson and Alexander Smith Clark for help with the \castro\ setup. The spectral 
data for the observed supernovae that we compared our simulation results to were obtained through WISeREP 
(\citealt{Yaron2012}; \url{https://www.wiserep.org}). This material is based in part upon work supported by the U.~S.\ 
Department of Energy, Office of Science, Office of Advanced Scientific Computing Research, Department of Energy 
Computational Science Graduate Fellowship under Award Number DE-SC0021110. DK is supported in part by the U.S. 
Department of Energy, Office of Science, Office of Nuclear Physics, DE-AC02-05CH11231, DE-SC0004658, and DE-SC0024388, 
and by a grant from the Simons Foundation (622817DK). This research benefited from meetings and collaborations funded 
by the Gordon and Betty Moore Foundation through Grant GBMF5076.  This research used resources of the National Energy 
Research Scientific Computing Center, a DOE Office of Science User Facility supported by the Office of Science of the 
U.~S.\ Department of Energy under Contract No.\ DE-AC02-05CH11231 using NERSC awards DDR-ERCAP 0032257 and NP-ERCAP 
0033809. 

This report was prepared as an account of work sponsored by an agency of the United States Government. Neither the United States Government nor any agency thereof, nor any of their employees, makes any warranty, express or implied, or assumes any legal liability or responsibility for the accuracy, completeness, or usefulness of any information, apparatus, product, or process disclosed, or represents that its use would not infringe privately owned rights. Reference herein to any specific commercial product, process, or service by trade name, trademark, manufacturer, or otherwise does not necessarily constitute or imply its endorsement, recommendation, or favoring by the United States Government or any agency thereof. The views and opinions of authors expressed herein do not necessarily state or reflect those of the United States Government or any agency thereof.

\section*{Data Availability}

The slightly-modified version of \castro\ used for our hydrodynamics simulations is available at
\url{https://github.com/KiranEiden/Castro/releases/tag/ce_sne_paper}, and the problem setup is
in the Exec/science/magnetar\_supernova subdirectory. The version of the
\textsc{Microphysics} repository used is available at
\url{https://github.com/KiranEiden/Microphysics/releases/tag/ce_sne_paper}. The final set of
simulations used version 24.01 of \textsc{AMReX} (\url{https://github.com/AMReX-Codes/amrex/releases/tag/24.01}). The 
spectral data for the observed supernovae that we compared our simulation results to can be accessed via the WISeREP 
archive (\citealt{Yaron2012}; \url{https://www.wiserep.org}).
Some numerical data from our simulation runs is available upon request.


\bibliographystyle{mnras}
\bibliography{refs}


\bsp	
\label{lastpage}
\end{document}